\documentclass[sigconf]{acmart}
\acmConference[ICSE 2026]{The 48th International Conference on Software Engineering}{April 12–18, 2026}{Rio de Janeiro, Brazil}
\usepackage{booktabs} 
\usepackage{multirow}
\usepackage{multicol}
\usepackage{enumerate}
\usepackage{tabularx} 
\usepackage{tabu}
\usepackage{threeparttable}
\usepackage{enumitem}
\usepackage{comment}
\usepackage{verbatim}
\usepackage[hyphenbreaks]{breakurl}
\usepackage{microtype}
\usepackage[utf8]{inputenc} 
\usepackage{balance}
\usepackage{siunitx}
\usepackage{tcolorbox}
\usepackage{graphicx} 
\usepackage{subfigure}
\usepackage{makecell}
\usepackage{diagbox}
\usepackage{soul}  
\usepackage{adjustbox}
\usepackage[figuresright]{rotating}
\usepackage{soul}

\newtcolorbox{summarybox}[1][]
{
    sharp corners,
    left=1mm,
    right=1mm,
    boxrule=0.3mm,
    colback=yellow!30!white
    #1,
}

\copyrightyear{2026}
\acmYear{2026}
\setcopyright{rightsretained}
\acmConference[ICSE '26]{48th International Conference on Software Engineering}{April 12–18, 2026}{Rio de Janeiro, Brazil}
\acmBooktitle{48th International Conference on Software Engineering (ICSE '26), April 12–18, 2026, Rio de Janeiro, Brazil}
\acmDOI{xxx}
\acmISBN{xxx}

\AtBeginDocument{%
  \providecommand\BibTeX{{%
    \normalfont B\kern-0.5em{\scshape i\kern-0.25em b}\kern-0.8em\TeX}}}

\begin{document}

\title{Evaluating Generated Commit Messages with Large Language Models}

\author{Qunhong Zeng}
\affiliation{
  \institution{Beijing Institute of Technology}
  \city{Beijing}
  \country{China}
}
\email{qunhongzeng@bit.edu.cn}

\author{Yuxia Zhang}\authornote{Corresponding authors}
\affiliation{
  \institution{Beijing Institute of Technology}
  \city{Beijing}
  \country{China}
}
\email{yuxiazh@bit.edu.cn}

\author{Zexiong Ma}
\affiliation{
  \institution{Peking University}
  \city{Beijing}
  \country{China}
}
\email{mazexiong@stu.pku.edu.cn}

\author{Bo Jiang}
\affiliation{
  \institution{ByteDance}
  \city{Shenzhen}
  \country{China}
}
\email{jiangbo.jacob@bytedance.com}

\author{Ningyuan Sun}
\affiliation{
  \institution{ByteDance}
  \city{Shenzhen}
  \country{China}
}
\email{sunningyuan@bytedance.com}

\author{Klaas-Jan Stol}
\affiliation{
  \institution{University College Cork}
  \city{Cork}
  \country{Ireland}
}
\email{k.stol@ucc.ie}

\author{Xingyu Mou}
\affiliation{
  \institution{Beijing Institute of Technology}
  \city{Beijing}
  \country{China}
}
\email{xingyum@bit.edu.cn}

\author{Hui Liu}
\affiliation{
  \institution{Beijing Institute of Technology}
  \city{Beijing}
  \country{China}
}
\email{liuhui08@bit.edu.cn}

\renewcommand{\shortauthors}{Zeng et al.}

\begin{abstract}
Commit messages are essential in software development as they serve to document and explain code changes. Yet, their quality often falls short in practice, with studies showing significant proportions of empty or inadequate messages. While automated commit message generation has advanced significantly, particularly with Large Language Models (LLMs), the evaluation of generated messages remains challenging. Traditional reference-based automatic metrics like BLEU, ROUGE-L, and METEOR have notable limitations in assessing commit message quality, as they assume a one-to-one mapping between code changes and commit messages, leading researchers to rely on resource-intensive human evaluation. This study investigates the potential of LLMs as automated evaluators for commit message quality. Through systematic experimentation with various prompt strategies and state-of-the-art LLMs, we demonstrate that LLMs combining Chain-of-Thought reasoning with few-shot demonstrations achieve near human-level evaluation proficiency. Our LLM-based evaluator significantly outperforms traditional metrics while maintaining acceptable reproducibility, robustness, and fairness levels despite some inherent variability. This work conducts a comprehensive preliminary study on using LLMs for commit message evaluation, offering a scalable alternative to human assessment while maintaining high-quality evaluation.
\end{abstract}



\maketitle

\section{Introduction}
\label{sec: intro}
Commit messages serve as a communication medium between developers to document code changes, and are integral to software development with version control systems \cite{buse2010automatically}. A high-quality commit message should contain two elements: the \textit{`What,'} a summary of what changed, and the \textit{`Why'}: the rationale behind these changes. These two elements help developers to quickly grasp code changes without having to dive too much into implementation details and the commit history to understand the reason behind specific code changes \cite{tian2022makes, li2023commit}.

Despite the generally acknowledged importance of clear commit messages, developers may fail to write high-quality messages because of tight deadlines or a lack of motivation \cite{nie2021coregen}. This mismatch between best practices and reality is underscored by prior findings that over 14\% of commit messages in SourceForge-hosted projects were empty \cite{dyer2013boa}, while nearly 44\% of commit messages in high-activity open source projects demanded significant quality improvements \cite{tian2022makes}. These gaps in commit message quality pose considerable risks to project sustainability, as incomplete or poorly written messages hinder effective collaboration and complicate code comprehension \cite{tian2022makes}. Consequently, researchers have invested substantial efforts in automating commit message generation (CMG) to reduce developers’ workload while improving the overall quality of commit messages.

Recently, significant advances have been made in automatic CMG \cite{Loyola_2017, loyola2018content, xu2019commit, liu2019generating, nie2021coregen, dong2022fira}, evolving from early rule-based approaches to leveraging Large Language Models (LLMs) \cite{diffisno, 10713474, zhang2024using}. Researchers typically evaluate the quality of these approaches using reference-based automatic metrics, including BLEU \cite{papineni2002bleu}, ROUGE-L \cite{lin2004rouge}, and METEOR \cite{banerjee-lavie-2005-meteor}. These reference-based metrics evaluate generated messages by computing their textual similarity with developer-written reference messages. However, this approach has three significant limitations. First, evaluation scores become unreliable when reference messages are of lower quality than generated ones. Second, these metrics capture \textit{textual} rather than \textit{semantic} similarities, failing to recognize messages that convey the same meaning but using different expressions. Third, different developers may write semantically distinct yet equally valid commit messages for the same diff. For example, to summarize changes in a commit, while some developers might compare implementation alternatives, others might focus on describing the functionality \cite{tian2022makes}. In such cases, generated messages might effectively describe code changes but receive low scores due to limited textual or semantic overlap with the developer-written reference message. These limitations have been acknowledged, and researchers have incorporated human evaluation to provide more comprehensive assessments \cite{diffisno, wanghard, dong2022fira, shi2022-race}. For instance, Li et al. \cite{diffisno} evaluated commit messages across four dimensions: Comprehensiveness (how good commit messages describe a summary of what has been changed), Rationality (how good commit messages provide a logical explanation for the code changes), Conciseness, and Expressiveness. They found that their approach scored lower than the baseline on BLEU and ROUGE-L metrics, but performed better in human evaluation. Again, this reveals that the widely used reference-based metrics may be unreliable in the evaluation of generated commit message.

Although human evaluation remains the gold standard for assessing commit message quality, human evaluation is time-consuming, resource-intensive, and therefore lacks scalability. Reproducibility is also challenging due to evaluator availability and consistency issues. Recent advances in generative AI technology \cite{Chiang2023CanLL, liu-etal-2023-g} have demonstrated the effectiveness of LLMs in evaluating natural language processing (NLP) tasks that require diversity and creativity, such as open-ended story generation \cite{Chiang2023CanLL} and dialogue generation \cite{liu-etal-2023-g}. These promising results suggest the potential of using LLMs to evaluate generated commit messages, thereby overcoming the limitations associated with human evaluation. To the best of our knowledge, how LLMs perform in evaluating commit messages, and whether they are comparable to human evaluators, remain open questions. Thus, this paper presents the results of the first empirical investigation of LLM-based evaluation of commit messages, guided by three research questions.

First, we investigated whether LLM-based evaluators can perform comparably to human evaluators in assessing commit messages. Thus, we ask \textbf{RQ1: Are LLM-based evaluators comparable to human evaluators in commit message evaluation?} 
We leverage an LLM-based evaluator to assess commit messages and use Spearman \cite{spearman1961proof} and Kendall \cite{kendall1938new} correlation coefficients to evaluate the correlation between the LLM-based evaluator and human developers. Apart from directly prompting LLMs to perform the commit message evaluation task, utilizing their zero-shot in-context learning capability through carefully constructed prompts that incorporate evaluation task descriptions, assessment criteria, and target examples, we also explore various prompting strategies to enhance LLMs' zero-shot learning performance. We systematically examine combinations of different prompting techniques, including Chain-of-Thought reasoning (providing structured evaluation steps) \cite{cot}, few-shot demonstrations \cite{touvron2023llama}, and alternative scoring approaches (unified multi-dimensional assessment within a single prompt versus separate evaluation of individual dimensions). Our experiments use multiple state-of-the-art LLMs, including proprietary model GPT-4 \cite{achiam2023gpt} and open-source ones (Llama 3.3 \cite{dubey2024llama}, Qwen 2.5 \cite{yang2024qwen2}, QwQ \cite{qwq}). The experimental results reveal that an LLM-based evaluator with an appropriate prompting strategy correlates highly with human evaluators, indicating comparable performance in commit message evaluation.

Since LLM-based evaluators exhibit comparable evaluation performance to human developers' assessments, it is of interest to explore how LLM-based evaluators compare with traditional automatic evaluation metrics commonly used in the CMG literature. Thus, we ask \textbf{RQ2: What is the performance of the LLM-based evaluator compared to automatic evaluation metrics?} We implemented the optimal prompt strategy from RQ1 as our LLM-based evaluator. Our comparison encompasses both established metrics from CMG literature (BLEU \cite{papineni2002bleu}, ROUGE-L \cite{lin2004rouge}, METEOR \cite{banerjee-lavie-2005-meteor}, CIDEr \cite{vedantam2015cider}) and widely-adopted NLP metrics previously unexplored in commit message generation, BERTScore \cite{zhang2019bertscore} and SBERT (Sentence BERT) \cite{reimers2019sentence}. Analysis of Spearman \cite{spearman1961proof} and Kendall \cite{kendall1938new} correlation coefficients reveals that the LLM-based evaluator outperforms traditional automatic metrics in commit message quality assessment by a large margin, demonstrating superior alignment with human evaluation. Through thematic analysis \cite{cruzes2011recommended}, we further investigate the underlying factors contributing to the limited correlation between commonly used reference-based metrics and human evaluation.

Despite demonstrating comparable performance to human evaluators and outperforming traditional widely-used metrics by a large margin, LLMs' non-deterministic behavior in commit message evaluation raises stability concerns. We sought to investigate three common concerns about reproducibility, robustness, and fairness. Thus, we ask \textbf{RQ3: Can the performance of the LLM-based evaluators keep stable?} Reproducibility examines evaluation stability for identical inputs by analyzing scoring consistency across multiple evaluations with controlled temperature settings. Robustness assesses stability across semantically equivalent commit messages with different expressions. Fairness investigates potential biases in LLM-based evaluators. While LLMs exhibit some scoring variability, they demonstrate acceptable reproducibility and robustness without significant discriminatory bias.

The main contributions of this work are as follows:
\begin{enumerate}
\item We present the first investigation of utilizing LLMs for commit message quality evaluation, exploring effective implementation strategies and demonstrating strong correlations with human developer assessments.
\item We demonstrate that the LLM-based evaluator significantly outperforms widely used reference-based automatic metrics.
\item We conduct a comprehensive stability analysis of LLM-based commit message evaluation, examining its reproducibility, robustness, and fairness.
\end{enumerate}

We provide a replication package to facilitate reproducibility and future work: \url{https://anonymous.4open.science/r/4k8asu}.

\section{Background}
\label{sec: background}
In this section, we first introduce the task and related works of commit message generation in Sec. \ref{sec: cmg}. Next, we discuss two widely used evaluation methodologies in commit message generation research: automatic metrics evaluation (Sec. \ref{sec: evaluation_metrics}) and human evaluation (Sec. \ref{sec: human_eval}). Finally, Sec. \ref{sec: llm} presents background on LLMs and techniques for enhancing their zero-shot learning capabilities.

\subsection{Commit Message Generation}
\label{sec: cmg}
Commit message generation (CMG) is an automated software engineering task that aims to generate descriptive and informative commit messages for code changes in version control systems. While most CMG approaches generate messages based primarily on code diffs (changes between two versions of code), some incorporate additional inputs such as related issues and pull requests. Existing CMG approaches can be categorized into four types: rule-based, retrieval-based, learning-based, and hybrid approaches \cite{zhang2024automatic}.

Rule-based approaches \cite{2010Automatically, cortes2014automatically, 2016shen} rely on predefined patterns to analyze code modifications. When changes match these patterns, corresponding template-based messages are generated. However, these methods have limited capability to explain the rationale behind changes and are inflexible if modifications can not be aligned with the predefined patterns.
Retrieval-based approaches \cite{liu2018neural, hoang2020cc2vec} identify similar code changes in existing datasets, then adopt their corresponding developer-written messages. Their effectiveness depends heavily on the similarity search algorithm. Moreover, even seemingly similar code changes may not justify identical commit messages when the underlying context differs.
Learning-based approaches \cite{jiang2017automatically, Loyola_2017, loyola2018content, xu2019commit, liu2019generating, nie2021coregen, dong2022fira} frame CMG as a machine translation task, employing various methods to represent code changes. For instance, CoreGen \cite{nie2021coregen} utilizes a transformer architecture to represent code changes as token sequences. FIRA \cite{dong2022fira} performs fine-grained graph representation by parsing the Abstract Syntax Tree for each change hunk, utilizing both edit operations and sub-tokens.
Finally, hybrid approaches \cite{diffisno, shi2022-race, wang2021context, liu2020atom} combine retrieval-based and learning-based methods. For example, RACE \cite{shi2022-race} retrieves similar commits based on code change similarity and incorporates both retrieved code changes and messages into the generation process. OMG \cite{diffisno} enhances generation by retrieving related Pull Request titles and issue reports as additional context, combining these with code changes and utilizing GPT4 for message generation, achieving state-of-the-art performance.

The quality of generated messages in prior studies is evaluated through two primary methods: automatic metrics evaluation and human evaluation. Table~\ref{tab:evaluation_metrics} summarizes these evaluation methods.

\begin{table}[!b]
\centering
\setlength{\tabcolsep}{2.2pt}

\caption{Overview of evaluation metrics used in commit message generation approaches}
\begin{tabular}{p{2.5cm}ccccc}
\toprule
\multirow{2}{*}{\textmd{Approach}} & \multicolumn{4}{c}{Automatic} & \multirow{2}{*}{\textmd{Human}} \\
\cmidrule(lr){2-5}
& \textmd{BLEU} & \textmd{ROUGE} & \textmd{METEOR} & \textmd{CIDEr} \\

\midrule
DeltaDoc \cite{2010Automatically} & & & & & \checkmark \\   
ChangeScribe \cite{cortes2014automatically} & & & & & \checkmark \\   
AutoSumCC \cite{2016shen} & & & & & \checkmark \\   
\midrule
NNGen \cite{liu2018neural} & \checkmark & & & & \checkmark \\
LogGen \cite{hoang2020cc2vec}  & \checkmark & & & &  \\
\midrule
CmtGen \cite{jiang2017automatically} & \checkmark & & & & \checkmark \\
NMT \cite{Loyola_2017} & \checkmark & & & \\
ContextNMT \cite{loyola2018content} & \checkmark & & \checkmark & & \checkmark \\
CODISUM \cite{xu2019commit} & \checkmark & & \checkmark  & \\
PtrGNCMsg \cite{liu2019generating} & \checkmark & \checkmark &  & \\
CoreGen \cite{nie2021coregen} & \checkmark & \checkmark & \checkmark & \\
FIRA \cite{dong2022fira} & \checkmark & \checkmark & \checkmark & & \checkmark \\
\midrule
RACE \cite{shi2022-race} & \checkmark & \checkmark & \checkmark & \checkmark & \checkmark \\
CoRec\cite{wang2021context} & \checkmark & \checkmark & \checkmark & & \checkmark \\
ATOM \cite{liu2020atom} & \checkmark & \checkmark & \checkmark & & \checkmark  \\
OMG \cite{diffisno} & \checkmark & \checkmark & \checkmark & & \checkmark \\
\bottomrule
\end{tabular}
\label{tab:evaluation_metrics}
\end{table}

\subsection{Automatic Metrics Evaluation for CMG}
\label{sec: evaluation_metrics}
Among prior CMG studies, four automatic metrics have been used to evaluate the quality of generated commit messages: BLEU \cite{papineni2002bleu}, ROUGE \cite{lin2004rouge}, METEOR \cite{banerjee-lavie-2005-meteor}, and CIDEr \cite{vedantam2015cider}. This section briefly introduces these metrics since they are widely used and the formulas have been well documented. We provide a detailed introduction in our online appendix \cite{appendix}.

\textbf{BLEU} is a widely adopted precision-based metric. It calculates the precision for different n-gram lengths (typically 1-gram to 4-gram) and then takes a weighted geometric average of these precision scores.

\textbf{ROUGE} comprises a series of variant metrics \cite{lin2004rouge}. The two most frequently used variants are ROUGE-N and ROUGE-L. ROUGE-N measures the n-gram recall between the generated and reference messages, whereas ROUGE-L computes the length of the longest common subsequence. ROUGE-L is widely used in commit message generation literature.

\textbf{METEOR} is an evaluation metric based on the harmonic mean of unigram precision and recall, with additional linguistic features like stemming, synonym matching, and word order \cite{banerjee-lavie-2005-meteor}. In calculating precision and recall, METEOR considers not only exact unigram matches but also stem matching and synonymy matching.

\textbf{CIDEr} is an evaluation metric designed initially for image caption evaluation \cite{vedantam2015cider}, which has been introduced to evaluate commit message quality by Shi et al. \cite{shi2022-race}. It measures the similarity between generated and reference commit messages using TF-IDF weighted n-gram matching.

\subsection{Human Evaluation for CMG}
\label{sec: human_eval}
As we argued in Sec. \ref{sec: intro}, automatic reference-based evaluation metrics may not adequately assess the quality of generated commit messages, because developers may write semantically distinct yet equally valid commit messages for the same diff. Consequently, several studies \cite{2010Automatically,cortes2014automatically,2016shen,liu2018neural,jiang2017automatically,loyola2018content,dong2022fira,shi2022-race,wang2021context,liu2020atom,diffisno} have incorporated human evaluation for further quality assessment. Early approaches such as NNGen \cite{liu2018neural}, ContextNMT \cite{loyola2018content}, and FIRA \cite{dong2022fira} used a 5-point scale for developers to rate the semantic similarity between generated messages and their references, while CmtGEN \cite{jiang2017automatically} adopted an 8-point scale, where higher scores indicate greater similarity to the reference messages.

Although human developers can understand semantic similarities between different messages, reference-based evaluation approaches have inherent limitations.
Developers may compose different yet equally valid messages for the same code changes \cite{tian2022makes}, making similarity to the reference insufficient for evaluation. Recent works have shifted toward multi-dimensional assessment using 5-point Likert scales without references. For instance, CoRec \cite{wang2021context} evaluates messages along three dimensions: \textit{relevance}, \textit{usefulness}, and \textit{adequacy}. Similarly, RACE \cite{shi2022-race} and HORDA \cite{10.1145/3695996} assess \textit{informativeness}, \textit{conciseness}, and \textit{expressiveness}, while OMG analyzed existing human evaluation metrics and consolidated them into four dimensions: \textit{comprehensiveness} (the effectiveness of describing code changes, i.e., \textit{What}), \textit{rationality} (the quality of explaining change rationale, i.e., \textit{Why}), \textit{conciseness}, and \textit{expressiveness} \cite{diffisno}.

Among these dimensions, \textit{conciseness} and \textit{expressiveness} focus on linguistic quality, where current generation methods demonstrate strong performance. For instance, Wang et al. \cite{10.1145/3695996} reports that RACE \cite{shi2022-race} achieves average scores of 4.86 and 4.82 out of 5 in these aspects, respectively. In light of these high scores, and considering that our study requires a larger annotation volume than \cite{10.1145/3695996}, we focus on content-oriented metrics and exclude language quality. According to Tian et al. \cite{tian2022makes}, commit message quality fundamentally comprises \textit{What} and \textit{Why}. The evaluation metrics of OMG \cite{diffisno} align with this definition of a good commit message, where \textit{comprehensiveness} and \textit{rationality} can be seen as specific sub-dimensions of \textit{informativeness}.
Therefore, this study’s human evaluation specifically targets these two fundamental aspects of commit message content quality: \textit{What} and \textit{Why}.

\subsection{Large Language Models}
\label{sec: llm}

Large Language Models (LLMs) harness billions of parameters and are pre-trained on extensive natural language corpora. While proprietary models like GPT-4 \cite{achiam2023gpt} initially dominated the field, recent open source alternatives such as the DeepSeek family \cite{liu2024deepseek}, Qwen series \cite{yang2024qwen2}, and LLaMA variants \cite{dubey2024llama} have demonstrated comparable capabilities, gradually narrowing the performance gap across various benchmarks such as coding. These models have been widely adopted in software engineering tasks, such as code generation \cite{jiang2024self} and also commit message generation \cite{diffisno}. Built on multi-layer transformer architectures \cite{vaswani2017attention}, these models have shown that scaling, either through increased model parameters or expanded training data, unlocks remarkable abilities in tackling complex tasks \cite{zhao2023survey}. LLMs exhibit exceptional performance on unseen tasks, allowing them to perform tasks given only instructions without task-specific training effectively; this kind of ability is called zero-shot in-context learning \cite{Chiang2023CanLL}.

Various techniques exist to enhance LLMs' zero-shot in-context learning performance. One common approach involves additional training or fine-tuning on target tasks, which requires appropriate available data and computational resources. Instead of further training, LLMs' in-context learning capabilities can be improved through prompting strategies, such as few-shot learning \cite{brown2020language} or chain of thought (CoT) reasoning \cite{wei2022chain}. Few-shot learning enhances model performance on specific tasks by providing a few input-output exemplars in the prompt. CoT simulates human problem-solving processes by incorporating step-by-step logical deductions. This approach includes not only the problem description in the prompt but also guides the model to generate intermediate reasoning steps, forming a logical chain from the problem to its solution to enhance target task performance.

In this study, we employ various state-of-the-art LLMs including GPT-4 \cite{achiam2023gpt}, Llama-3.3-70B-Instruct \cite{dubey2024llama}, Qwen2.5-72B-Instruct \cite{yang2024qwen2}, and QwQ-32B \cite{qwq}. Among these, QwQ-32B is a lightweight reasoning model (comprising only 32 billion parameters) similar to DeepSeek-R1 \cite{guo2025deepseek}, which is trained to provide a reasoning process before delivering task answers to improve problem-solving performance. We test these models with different prompting strategy combinations to evaluate their capabilities in assessing commit message quality.

\section{Dataset} \label{sec: dataset}

We constructed a benchmark dataset comprising code changes, developer-written commit messages, and corresponding generated commit messages derived from the code changes. For each commit, both the developer-written and generated messages were evaluated by human developers. This section describes our data selection and human evaluation protocol.

\subsection{Data Preparation}

\subsubsection{Sample}
Given the resource-intensive nature of human evaluation and the complexity of recruiting annotators proficient in multiple programming languages, we sought to balance data diversity with feasibility by focusing on two widely used languages: Python and Java. Our data source is the CommitBench dataset \cite{schall2024commitbench}, which considers redistribution licenses and enforces various quality filters, including removing bot-generated messages and privacy-sensitive content.
We randomly sampled 200 Python commits and 200 Java commits from CommitBench\_long\footnote{\url{https://huggingface.co/datasets/Maxscha/commitbench_long}}, which uses an extended token limit of 2,048 instead of the default 512\footnote{\url{https://huggingface.co/datasets/Maxscha/commitbench}}, as Zhang et al. \cite{zhang2024automatic} found a median count of 632 tokens in diffs. Since our evaluation mainly focuses on commit message content, we excluded auxiliary information such as issue links and other URLs. While CommitBench attempts to sanitize issue references by replacing GitHub issue tags with the label \textit{\textless I\textgreater}, this approach is limited to GitHub's issue format and does not address references from other issue tracking platforms or custom issue tags. Additionally, some personal names or other privacy-sensitive data remained in the dataset.
To this end, we manually reviewed and processed each sampled developer-written message by:
\begin{enumerate}
    \item Removing issue tags (both CommitBench's \textit{\textless I\textgreater} replacements and unprocessed custom issue tags) and URLs.
    \item Eliminating personal names and email addresses.
    \item Excluding revert commits (replacing them with newly sampled non-revert commits).
\end{enumerate}

\subsubsection{Generate Messages for the Sampled Commits}
To address RQ2, we aim to compare LLM-based evaluators with commonly used automatic reference-based metrics, such as METEOR and ROUGE. Thus, we need to prepare generation messages for the 400 selected commits. 
To ensure a comprehensive evaluation of the metrics' ability to discriminate messages with diverse qualities, we intentionally introduced quality variation in the generated messages. Previous work that used LLMs for commit message generation \cite{diffisno, 10713474, zhang2024using} shows that models of different sizes can exhibit varying capabilities. Therefore, we employed three LLMs of different scales: GPT-4 \cite{achiam2023gpt}, Llama-3.3-70B-Instruct \cite{dubey2024llama}, and Qwen-2.5-14B-Instruct \cite{yang2024qwen2}, with each model generating one-third of the commit messages. However, our preliminary annotation results showed that most LLM-generated messages received high scores ($\ge3$) on the What dimension. To ensure a wide quality range in our benchmark, we deliberately instructed the models to generate poor-quality commit messages for 20\% of randomly selected samples. Details of the implementation and prompts are available in the appendix \cite{appendix}.

\subsection{Human Evaluation}
\label{sec:dataset_human_label}

The human evaluation focuses on two fundamental dimensions of commit message content quality: \textit{What} and \textit{Why}. The definitions of these two dimensions are \cite{tian2022makes, diffisno}:
\begin{enumerate}
\item \textbf{What}: How well the commit message accurately captures the changes made in the code.
\item \textbf{Why}: How well the commit message explains the rationale behind making the change.
\end{enumerate}

We recruited six volunteers, each with over five years of software development experience and Python and Java programming proficiency. From our dataset of 400 commits, each code change was associated with two commit messages: one written by the original developer and one generated by an LLM. Similar to previous work \cite{loyola2018content,dong2022fira, shi2022-race,diffisno}, we used a 5-point Likert scale (0 for poor, 1 for marginal, 2 for acceptable, 3 for good, and 4 for excellent). To ensure the reliability and consistency of the evaluation, we conducted a three-stage annotation process. First, in the pilot annotation stage, each volunteer independently assessed 50 identical commit samples. The volunteers were granted access to the commit's git repository to facilitate a good understanding of the commits under evaluation, scoring both the \textit{What} and \textit{Why} dimensions for each commit message. This stage aimed to familiarize the volunteers with the evaluation task and establish an initial understanding of scoring criteria. Second, after the pilot annotation stage, we conducted a discussion and calibration session where the volunteers explained the reasons behind their different scores for the same commit message. This stage allowed the volunteers to learn from each other, understand different perspectives, and reach a consensus on the scoring criteria. Finally, in the formal evaluation stage, the volunteers independently assessed the remaining 750 commit messages. To minimize potential bias, we randomly shuffled the evaluation samples. Each volunteer received randomly shuffled commits (code changes, messages, and git repository information) for evaluation, without knowing whether the commit messages were human-written or LLM-generated.

\begin{figure}[!h]
  \centering
  \includegraphics[width=0.33\textwidth]{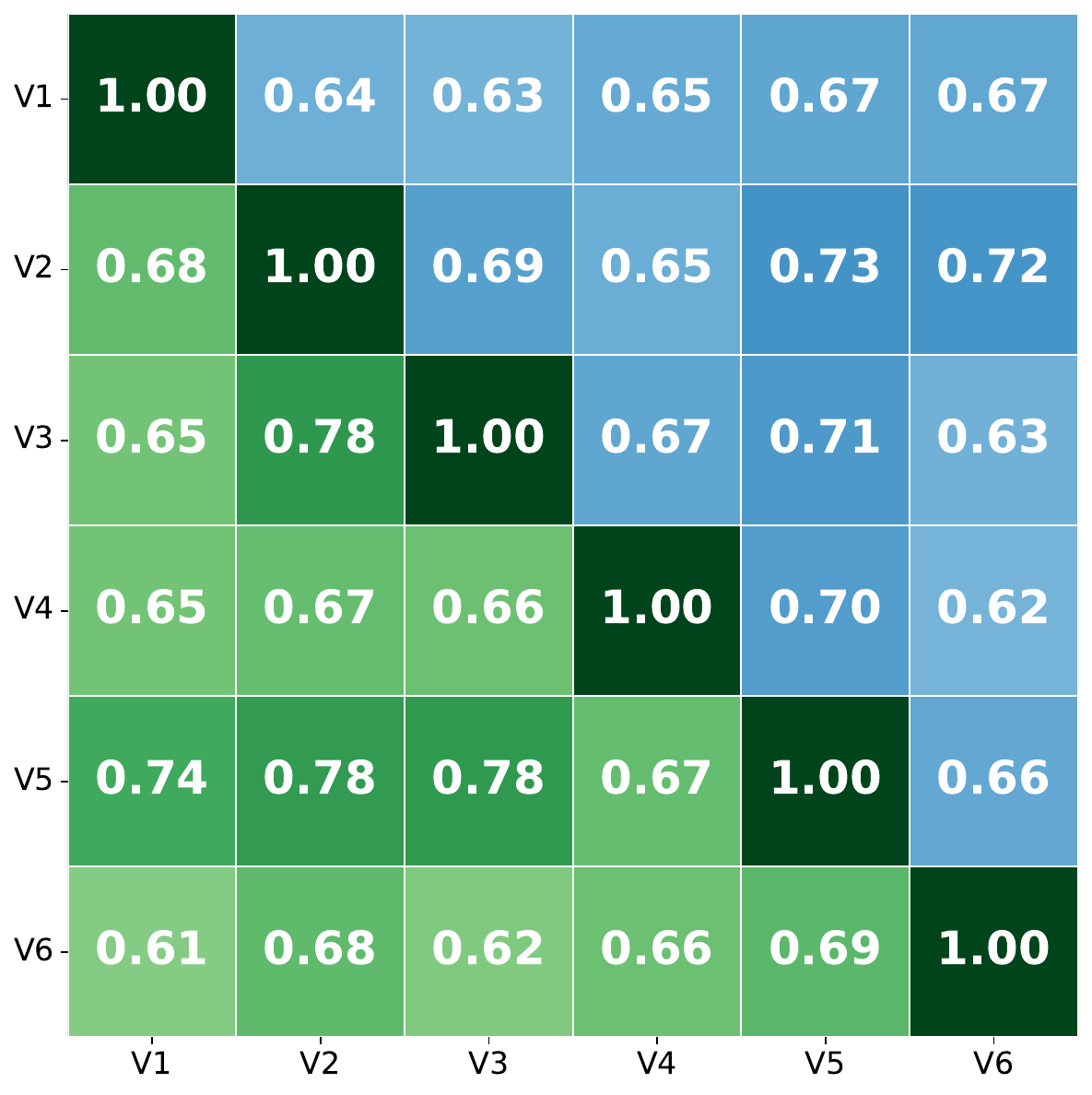}
  \caption{Spearman Correlation Coefficient of Volunteers, the upper triangle (blue) represents What, the lower triangle (green) represents Why}
  \label{fig: spearman_of_volunteers}
\end{figure}

To assess the inter-rater reliability of the six volunteers for both the \textit{What} and \textit{Why} dimensions, we calculated Spearman correlation coefficients \cite{spearman1961proof} between each pair of volunteers' scores. The results are shown in Figure \ref{fig: spearman_of_volunteers}. All correlation coefficients exceeded 0.6, indicating a high agreement \cite{kuckartz2013statistik} among volunteers. The final \textit{What} and \textit{Why} scores for each commit message were computed by averaging all volunteers' rating scores.

\subsection{Implementation Details}

We used various LLMs in our experiments. For the proprietary model GPT-4 (gpt-4-0613), we utilized the Azure API to send requests. For open source models, we deployed the backend using vLLM \cite{kwon2023efficient} on a cluster of 8x NVIDIA A100 40GB GPUs. Based on empirical performance evaluation of different temperature settings (detailed in Sec. \ref{sec: reproducibility}), we set the temperature parameter at 0.25 for proprietary models and 0 for open-source models. The top\_p parameter remained at its default value of 0.5 across all experiments. To increase reliability, final LLM scores for all experiments were derived by averaging three repeated evaluations of identical inputs. For automatic metrics implementation, we used the Evaluate library \cite{hf_evaluate} from the Hugging Face community to ensure implementation reliability. We implemented BertScore using the BertScore library \cite{bert-score} with its default roberta-large model, and Sentence-BERT Score using the sentence-transformers library with the all-MiniLM-L6-v2 model.\footnote{\url{https://huggingface.co/sentence-transformers/all-MiniLM-L6-v2}.} The CIDEr metric was implemented using the code provided by Shi et al. \cite{shi2022-race}.

\section{RQ1: Are LLM-based evaluators comparable to human evaluators in commit message evaluation?}
\label{sec: rq1}

\subsection{Method}

To evaluate commit message quality using LLMs, we first employed zero-shot in-context learning, assessing the \textit{What} and \textit{Why} dimensions using a 5-point Likert scale. The prompt provided to the LLM is: \textit{``You are an experienced software engineer. Your task is to evaluate the commit message based on the corresponding git commit diff, focusing on two dimensions: <dimensional definitions> + <scoring criteria> + <commit to be evaluated>.''} Within this prompt, we included explicit definitions of each dimension, detailed scoring criteria, and both the commit message under evaluation and its corresponding code changes.

Building upon the introduction of LLMs in Sec. \ref{sec: llm}, we then investigated three prominent strategies to enhance zero-shot in-context learning performance: Chain of Thought \cite{wei2022chain}, Few-Shot Learning \cite{brown2020language}, and their combination. In the Chain of Thought approach, we instructed the LLM to evaluate commit messages step-by-step. In the Few-Shot Learning approach, we provided the LLM with exemplary high- and low-quality examples to illustrate the evaluation criteria clearly. Due to space constraints, detailed descriptions of all prompts are provided in the online appendix \cite{appendix}. Additionally, we explored two scoring methodologies: unified evaluation, in which the LLM assesses both dimensions within a single prompt, and separate evaluation, in which each dimension is assessed individually. The aim was to identify the strategy most closely aligned with human evaluations.

We evaluated four state-of-the-art LLMs: GPT-4 \cite{achiam2023gpt}, Llama-3.3-70B-Instruct \cite{dubey2024llama}, Qwen2.5-72B-Instruct \cite{yang2024qwen2}, and QwQ-32B \cite{qwq}. Each LLM was tested using various combinations of prompting strategies to determine the optimal model and prompt configuration. We use 400 human-written commit messages from the dataset constructed in Section~\ref{sec: dataset}. For each combination of LLM and prompting strategy, suitable assessment prompts were constructed. To quantify the alignment between LLM-generated and human evaluator ratings, we calculated Spearman \cite{spearman1961proof} and Kendall \cite{kendall1938new} correlation coefficients, which are appropriate measures for paired data, following established methodologies from recent studies \cite{liu-etal-2023-g, Chiang2023CanLL, hu2022correlating}.

\begin{table*}[h]
\centering
\caption{Correlation between LLM-generated and Human Evaluation Scores}
\setlength{\tabcolsep}{5pt} 
\small
\label{tab:rq1}
\begin{tabular}{p{3cm}p{2cm}cccccccc}
\toprule
\multirow{2}{*}{\textbf{Model}} & \multirow{2}{*}{\textbf{Strategy}} & \multicolumn{2}{c}{\textbf{What  (Separate)}} & \multicolumn{2}{c}{\textbf{Why (Separate)}} & \multicolumn{2}{c}{\textbf{What (Unified)}} & \multicolumn{2}{c}{\textbf{Why (Unified)}} \\
\cmidrule(lr){3-4} \cmidrule(lr){5-6} \cmidrule(lr){7-8} \cmidrule(lr){9-10}
& & Spearman & Kendall & Spearman & Kendall & Spearman & Kendall & Spearman & Kendall \\
\midrule
\multirow{4}{*}{GPT4} 
& Zero-shot & 0.61 & 0.49 & 0.58 & 0.47 & \textbf{0.66} & \textbf{0.54} & 0.64 & 0.52 \\
& CoT & 0.62 & 0.48 & 0.70 & 0.56 & 0.65 & 0.52 & \textbf{0.78} & \textbf{0.66} \\
& Few-shot & \textbf{0.63} & \textbf{0.51} & 0.65 & 0.53 & 0.63 & 0.52 & 0.68 & 0.55 \\
& Zero-shot+CoT & \textbf{0.63} & 0.50 & \textbf{0.72} & \textbf{0.57} & 0.65 & 0.52 & \textbf{0.78} & 0.64 \\
\midrule
\multirow{4}{*}{QwQ-32B} 
& Zero-shot & 0.53 & 0.41 & 0.57 & 0.45 & 0.59 & 0.46 & 0.70 & 0.57 \\
& CoT & 0.54 & 0.42 & 0.62 & 0.49 & 0.60 & 0.46 & 0.72 & 0.60 \\
& Few-shot & 0.57 & 0.46 & 0.63 & \textbf{0.50} & \textbf{0.63} & \textbf{0.51} & \textbf{0.75} & \textbf{0.62} \\
& Zero-shot+CoT & \textbf{0.59} & \textbf{0.47} & \textbf{0.64} & \textbf{0.50} & 0.60 & 0.48 & 0.74 & 0.61 \\
\midrule
\multirow{4}{*}{Llama-3.3-70B-Instruct} 
& Zero-shot & 0.61 & 0.50 & 0.55 & 0.45 & 0.61 & 0.50 & 0.66 & 0.55 \\
& CoT & 0.55 & 0.44 & \textbf{0.64} & \textbf{0.51} & 0.56 & 0.44 & \textbf{0.74} & \textbf{0.60} \\
& Few-shot & \textbf{0.62} & \textbf{0.51} & 0.56 & 0.45 & \textbf{0.62} & \textbf{0.52} & 0.68 & 0.56 \\
& Zero-shot+CoT & 0.61 & 0.49 & 0.63 & 0.49 & 0.57 & 0.46 & 0.71 & 0.57 \\
\midrule
\multirow{4}{*}{Qwen2.5-72B-Instruct} 
& Zero-shot & \textbf{0.66} & 0.55 & 0.55 & 0.45 & \textbf{0.63} & \textbf{0.53} & 0.56 & 0.46 \\
& CoT & 0.58 & 0.48 & \textbf{0.60} & \textbf{0.48} & 0.61 & 0.50 & 0.60 & 0.49 \\
& Few-shot & 0.66 & \textbf{0.55} & 0.56 & 0.46 & 0.61 & 0.51 & 0.62 & 0.51 \\
& Zero-shot+CoT & 0.60 & 0.50 & 0.59 & \textbf{0.48} & 0.59 & 0.49 & \textbf{0.67} & \textbf{0.54} \\
\bottomrule
\end{tabular}
\end{table*}

\subsection{Results}
\label{sec: rq1_result}

Table \ref{tab:rq1} presents the results, where ``separate'' indicates separate evaluations of the \textit{What} and \textit{Why} dimensions, while ``unified'' represents a single evaluation process assessing both dimensions simultaneously. Most models demonstrated significant improvement in performance under unified evaluation compared to separate evaluation in the \textit{Why} dimension, with minor improvements or slight fluctuations in the \textit{What} dimension.

Our analysis of different prompting strategies revealed several findings. First, the Chain of Thought (CoT) strategy effectively improved correlation with human judgment in the \textit{Why} dimension across both evaluation approaches. For instance, GPT-4's Spearman correlation increased from 0.58 to 0.70 in separate evaluation and from 0.64 to 0.78 in unified evaluation. However, the CoT improvement in the \textit{What} dimension was limited, with some models showing slight improvements while others exhibited minor decreases. Notably, QwQ-32B maintained nearly unchanged performance between zero-shot and CoT conditions, as this reasoning model inherently outputs a reasoning process (CoT) regardless of whether CoT prompts are explicitly provided. The Few-shot strategy demonstrated consistent effectiveness across both dimensions, with all models benefiting from or maintaining performance levels in both evaluation settings.

When combining CoT and Few-shot strategies, most models maintained the improvements in the \textit{Why} dimension achieved through CoT, while performance in the \textit{What} dimension varied slightly among models, generally remaining at comparable levels. Based on these results, we selected GPT-4 with the CoT+Few-shot strategy using unified evaluation as our optimal configuration, achieving Spearman correlations of 0.65 and 0.78 for the \textit{What} and \textit{Why} dimensions, respectively. Notably, these correlation coefficients closely match the inter-rater correlations among human developer volunteers (see Figure \ref{fig: spearman_of_volunteers}), suggesting that GPT-4 with appropriate enhancements can achieve human-level proficiency in commit message evaluation.

\begin{summarybox}
{
\textbf{Summary for RQ1:} LLMs demonstrate comparable evaluation performance to human developers when evaluating commit messages across the dimensions of \textit{What} and \textit{Why}. Among the four LLMs of varying sizes and types evaluated in our study, even the lowest-performing model achieved a Spearman correlation of approximately .53, indicating a moderate to strong correlation. The correlation between LLM and human developer ratings can be further enhanced by selecting appropriate base models and implementing suitable prompting strategies. The best-performing model achieved Spearman correlations of .65 for the \textit{What} dimension and .78 for the \textit{Why} dimension, demonstrating human-level proficiency in commit message evaluation.
}
\end{summarybox}

\section{RQ2: What is the performance of the LLM-based evaluator compared to automatic evaluation metrics?}
\label{sec: rq2}
\subsection{Method}
This research question asks how well LLM-based evaluators perform when assessing generated commit messages, and how they compare to traditional, commonly used automatic evaluation metrics. To conduct a comprehensive comparison of evaluation metrics, we incorporated all metrics previously used in commit message generation studies, including BLEU, ROUGE, METEOR, and CIDEr (see Sec.~\ref{sec: evaluation_metrics} for their underlying principles and computational methods). Additionally, we introduced two semantic-based metrics, i.e., BERTScore \cite{zhang2019bertscore} and SBERT (Sentence BERT) \cite{reimers2019sentence}, which are not previously used in commit message evaluation but widely adopted in NLP tasks.

BERTScore is a text evaluation metric that measures semantic similarity by computing token-level cosine similarities using BERT embeddings \cite{zhang2019bertscore}. It creates a similarity matrix between candidate and reference texts, then uses greedy matching to calculate precision (matching candidate tokens to reference) and recall (matching reference tokens to the candidate). The final score is the F1 measure of these matches, enabling a more robust evaluation than traditional string-matching metrics.
Sentence-BERT (SBERT) \cite{reimers2019sentence} is a variant of BERT that creates sentence embeddings by converting text into fixed-length vectors. It processes each sentence independently through BERT and uses pooling to combine token embeddings into a single sentence vector. The similarity between two texts can be easily calculated using cosine similarity between their vectors.

Our evaluation involved 400 generated commit messages (see Sec. \ref{sec:dataset_human_label}), assessed using both LLM-based and automatic metrics. We used GPT-4 with CoT+Few-shot prompting strategy under unified evaluation (see Sec. \ref{sec: rq1_result}). Following established practices in commit message generation research, we used developer-written messages as references for automatic metrics. To assess metric alignment with human judgments, we calculated Spearman and Kendall correlation coefficients, consistent with our approach for addressing RQ1.

\subsection{Results}
\label{sec: rq2_result}

\begin{table}[h]
\robustify\bfseries
\centering
\small
\sisetup{
    group-digits=false,
    group-minimum-digits=4,
    table-format=0.3,
    mode=text,
    round-mode=places,
    round-precision=2,
    detect-weight=true, 
    detect-family=true
}
\caption{Correlation between Different Evaluation Metrics and Human Judgments}
\label{tab:metrics_comparison}
\begin{tabular}{p{2cm}
                S
                S
                S
                S}
\toprule
\multirow{2}{*}{\textbf{Metrics}} & \multicolumn{2}{c}{\textbf{What Dimension}} & \multicolumn{2}{c}{\textbf{Why Dimension}} \\
\cmidrule(lr){2-3}\cmidrule(lr){4-5}
& {Spearman} & {Kendall} & {Spearman} & {Kendall} \\
\midrule
\multicolumn{5}{l}{\textit{Textual-Similarity Metrics}} \\
\midrule
BLEU & 0.08 & 0.07 & 0.10 & 0.09 \\
ROUGE-L & 0.28 & 0.20 & 0.07 & 0.05 \\
METEOR & \bfseries 0.42 & \bfseries 0.31 & \bfseries 0.20 & \bfseries 0.14 \\
CIDEr & 0.24 & 0.17 & 0.09 & 0.06 \\
\midrule
\multicolumn{5}{l}{\textit{Semantic-Similarity Metrics}} \\
\midrule
BERTScore & 0.21 & 0.15 & 0.12 & 0.08 \\
SBERT & 0.34 & 0.25 & 0.18 & 0.13 \\
\midrule
\multicolumn{5}{l}{\textit{LLM-based Metrics}} \\
\midrule
GPT-4 & \bfseries 0.65 & \bfseries 0.56 & \bfseries 0.78 & \bfseries 0.63 \\
\bottomrule
\end{tabular}
\end{table}

Table \ref{tab:metrics_comparison} presents the evaluation results, with the best-performing metrics highlighted in bold. The results show that widely used textual and semantic similarity metrics perform poorly on the \textit{Why} dimension, with the highest Spearman correlation coefficient being only .20 and the highest Kendall correlation coefficient being .14, indicating a weak correlation with human developers' judgments \cite{kuckartz2013statistik, spearman1961proof}.
In the \textit{What} dimension, BLEU exhibits the weakest correlation with human assessments, attributable to its design as a corpus-level metric, which compromises reliability in sentence-level evaluations of generated commit messages \cite{papineni2002bleu}. Nonetheless, we include BLEU results for reference. Of the commonly used metrics, METEOR achieves superior performance with Spearman and Kendall correlations of .42 and .31, followed by SBERT with correlations of .34 and .25. Despite improved performance in the \textit{What} dimension, even METEOR demonstrates only moderate correlation with human judgments \cite{kuckartz2013statistik, spearman1961proof}. In contrast, LLM-based metrics, designed for distinct dimensional evaluation, achieve robust correlations in both dimensions when assessing generated commit messages, with Spearman correlations reaching .65 for \textit{What} and .78 for \textit{Why}, substantially surpassing all automatic metrics.

To further investigate the performance gap between LLM-based and reference-based automatic metrics in commit message evaluation, we conducted a thematic analysis  \cite{cruzes2011recommended}, focusing on METEOR, the best-performing automatic metric. Specifically, our analysis examined discrepancies between human and METEOR evaluations, identifying 101 examples that received high human ratings (top 30\%) for the \textit{What} dimension but low METEOR scores (bottom 30\%), and conversely, 1 example with a high METEOR score (top 30\%) but low human rating (bottom 30\%). Notably, we did not find any similar discrepancies with LLM metrics under the same filtering threshold, revealing the performance advantage of LLM metrics. We only considered the \textit{What} dimension because the correlation between all reference-based automatic metrics and human developers is low for the \textit{Why} dimension. Through a thematic analysis of these discrepancies and examining METEOR's computational principles (Sec. \ref{sec: evaluation_metrics}), we investigated why METEOR assigned low scores to commit messages that human evaluators rated highly. We identified four main categories:\footnote{Note that the sum of the instances in the four identified categories does not equal 102 because some instances belong to multiple categories.} Low-quality References, High-Expectation References, Lexical and Stylistic Variability, and Semantic Divergence. We discuss these next.

\textbf{Category 1: Low-quality References (\#32)} refers to vague or incomplete developer-written messages (e.g., \textit{``fix a bug''}), which exist in our dataset due to random sampling. In these cases, generated messages often surpass reference quality but share minimal vocabulary, resulting in low recall and, consequently, low METEOR scores despite high human ratings. This category is a common drawback for reference-based metrics. To address this issue, we selected a subset of the dataset where each reference's \textit{What} dimension human score was greater than or equal to 2.5, indicating a high-quality score. This filtering process left 236 cases; Table \ref{tab:metrics_comparison_on_high_quality} presents the correlation comparison. The results show that the correlation between reference-based metrics and human judgments improved for the high-quality sub-dataset. METEOR remains the best-performing reference-based metric; however, its correlation still lags behind LLM-based metrics.

\begin{table}[h]
\centering
\small
\caption{Correlation between Different Evaluation Metrics and Human Judgments on High-quality Subset}
\label{tab:metrics_comparison_on_high_quality}
\begin{tabular}{p{2cm}cccc}
\toprule
\multirow{2}{*}{\textbf{Metrics}} & \multicolumn{2}{c}{\textbf{What Dimension}} & \multicolumn{2}{c}{\textbf{Why Dimension}} \\
\cmidrule(lr){2-3}\cmidrule(lr){4-5}
& Spearman & Kendall & Spearman & Kendall \\
\midrule
\multicolumn{5}{l}{\textit{Textual-Similarity Metrics}} \\
\midrule
BLEU & 0.09 & 0.08 & 0.12 & 0.10 \\
ROUGE-L & 0.35 & 0.25 & 0.09 & 0.06 \\
METEOR & \bfseries 0.53 & \bfseries 0.38 & \bfseries 0.27 & \bfseries 0.19 \\
CIDEr & 0.31 & 0.22 & 0.12 & 0.08 \\
\midrule
\multicolumn{5}{l}{\textit{Semantic-Similarity Metrics}} \\
\midrule
BERTScore & 0.31 & 0.22 & 0.14 & 0.10 \\
SBERT & 0.40 & 0.29 & 0.16 & 0.11 \\
\midrule
\multicolumn{5}{l}{\textit{LLM-based Metrics}} \\
\midrule
GPT-4 & \bfseries 0.67 & \bfseries 0.58 & \bfseries 0.78 & \bfseries 0.64 \\
\bottomrule
\end{tabular}
\end{table}

\textbf{Category 2: High-Expectation References (\#11)} are messages that receive high scores from human developers and contain not only \textit{What} but also \textit{Why} information, as well as details that developers consider necessary to help readers understand the code changes. In such cases, the reference token count increases, resulting in a large recall denominator, requiring the generated message to include the same details. Although such references are the ultimate goal of automatic commit message generation, a generated message that does not include the same \textit{Why} information and details as the reference does not necessarily indicate a low-quality message in terms of the \textit{What} dimension.

\textbf{Category 3: Lexical and Stylistic Variability (\#22)} refers to cases where the generated and developer messages express the same semantics. However, there can be various ways to express the same meaning. Because of the expression diversity, the two messages may have little word overlap, leading to low recall and, consequently, a low METEOR score. For example, one reference message of a code change in our analyzed cases is \textit{``[Tiny]Remove duplicated assignment,''} while the generated message is \textit{``Remove redundant GcsClient instantiation.''} Although the two messages express similar semantics, they only share one matching word: \textit{``Remove.''}

\textbf{Category 4: Semantic Divergence (\#45)} is the most significant and fundamental reason why reference-based metrics fail to evaluate commit messages effectively. Essentially, the mapping from code changes to commit messages is not one-to-one. As identified by Tian et al. \cite{tian2022makes}, the \textit{What} and \textit{Why} dimensions can be expressed through various perspectives. For example, in one of our analyzed cases, the reference commit message, \textit{``Added hash\_create option so hashes can create new threads,''} describes the modification from a functional perspective. The corresponding generated commit message, \textit{``Add hash\_create parameter to add\_comment and update process\_hashes call,''} summarizes the change in terms of the modified code objects. These different but valid perspectives create semantic divergence, resulting in minimal lexical and semantic similarity despite both messages can be high-quality.

\begin{summarybox}
{
\textbf{Summary for RQ2:} 
LLM-based metrics significantly outperform traditional reference-based automatic metrics in assessing generated commit messages. The thematic analysis of examples that received high human but low METEOR scores reveals four categories explaining why reference-based metrics perform poorly in evaluating commit messages. The primary reason is the one-to-many nature of mapping code changes to messages, and using a single reference to calculate similarity is hindered by this diversity.
}
\end{summarybox}

\section{RQ3: Can the performance of the LLM-based evaluators keep stable?}
\label{sec: rq3}
Although LLM-based evaluators demonstrate superior correlation with human developer judgments, their non-deterministic nature may affect their performance stability. This RQ examines three critical aspects of LLM stability: reproducibility, robustness, and fairness. 

\subsection{Reproducibility}
\label{sec: reproducibility}
Reproducibility assesses the consistency of LLM-based evaluator scores across multiple evaluations of identical inputs (code changes and commit messages).

\subsubsection{Method}
The non-determinism of LLM-generated output stems from multiple sources, including system-level variations in hardware architectures and computing environments (e.g., non-deterministic CUDA operators). However, the primary source is the LLM's sampling mechanism, which typically employs probabilistic token selection based on predicted logits, controlled by the temperature hyperparameter \cite{brown2020language}. Temperature values (0-1) regulate token selection probability: lower values increase determinism by amplifying high-logit token probability, while higher values enhance randomness. To evaluate the temperature's impact on evaluator reproducibility, we assessed 400 developer-written messages with five temperature settings (0, 0.25, 0.5, 0.75, and 1), conducting three evaluations per message at each setting. We tested four models from RQ1, all using the CoT+few-shot prompting strategy under unified evaluation. For each message at each temperature setting, we calculated the standard deviation (std) and mean absolute difference (MAD) across the three evaluations. We then averaged these metrics across all 400 messages to assess the overall evaluation consistency at each temperature value.

\subsubsection{Results}
The results in Table \ref{tab:reproducibility} demonstrate distinct stability patterns across LLM evaluators. Open-source models (QwQ, Qwen2.5, and Llama3.3) deployed via VLLM \cite{kwon2023efficient} exhibited progressively decreasing evaluation stability with increasing temperature values, while the proprietary model GPT-4 accessed through its API demonstrated comparable levels of variation across all temperature settings, with optimal performance at temperature 0.25.
Taking GPT-4 at temperature 0.25 as an example, we analyzed 1,200 pairs of evaluations (derived from 400 triplets). In the \textit{What} dimension (mean std=0.24), 906 (75.5\%) pairs showed identical scores, with 258 pairs differing by one point and 36 pairs showing larger differences. The \textit{Why} dimension (mean std=0.29) exhibited similar patterns, with 890 (74.2\%) identical pairs, 215 one-point differences, and 95 pairs showing larger differences. Considering that a one-point scoring difference is generally acceptable (given that human developers may also face similar granularity in distinguishing between scores like 2 and 3), GPT-4 demonstrated stability in 97\% of cases for the \textit{What} dimension and 92\% for the \textit{Why} dimension. These results indicate that, in most cases, LLM-based evaluators can give stable scores for identical inputs. In scenarios that require high precision, we recommend using averaged scores from multiple repeated evaluations when employing LLMs for commit message assessment to increase result stability.

\begin{table}[!h]
\caption{Reproducibility analysis of different LLMs across temperature settings}
\label{tab:reproducibility}
\small
\centering
\begin{tabular}{llllccccc}
\toprule
\multirow{2}{*}{Model} & \multirow{2}{*}{Dim} & \multirow{2}{*}{Metric} & \multicolumn{5}{c}{Temperature} \\
\cmidrule(lr){4-8}
& & & 0.00 & 0.25 & 0.50 & 0.75 & 1.00 \\
\midrule
\multirow{4}{*}{GPT-4} & \multirow{2}{*}{What} & Std. & 0.29 & 0.24 & 0.26 & 0.26 & 0.28 \\
& & MAD & 0.34 & 0.28 & 0.31 & 0.31 & 0.33 \\
\cmidrule(lr){2-8}
& \multirow{2}{*}{Why} & Std. & 0.38 & 0.29 & 0.36 & 0.34 & 0.38 \\
& & MAD & 0.46 & 0.35 & 0.43 & 0.40 & 0.45 \\
\midrule
\multirow{4}{*}{QwQ} & \multirow{2}{*}{What} & Std. & 0.23 & 0.24 & 0.25 & 0.23 & 0.26 \\
& & MAD & 0.27 & 0.28 & 0.29 & 0.27 & 0.31 \\
\cmidrule(lr){2-8}
& \multirow{2}{*}{Why} & Std. & 0.27 & 0.29 & 0.29 & 0.28 & 0.31 \\
& & MAD & 0.32 & 0.34 & 0.35 & 0.33 & 0.37 \\
\midrule
\multirow{4}{*}{Llama-3.3} & \multirow{2}{*}{What} & Std. & 0.09 & 0.11 & 0.15 & 0.16 & 0.18 \\
& & MAD & 0.10 & 0.13 & 0.18 & 0.19 & 0.21 \\
\cmidrule(lr){2-8}
& \multirow{2}{*}{Why} & Std. & 0.24 & 0.27 & 0.29 & 0.32 & 0.31 \\
& & MAD & 0.28 & 0.32 & 0.34 & 0.38 & 0.37 \\
\midrule
\multirow{4}{*}{Qwen2.5} & \multirow{2}{*}{What} & Std. & 0.01 & 0.06 & 0.09 & 0.10 & 0.14 \\
& & MAD & 0.01 & 0.07 & 0.11 & 0.11 & 0.17 \\
\cmidrule(lr){2-8}
& \multirow{2}{*}{Why} & Std. & 0.02 & 0.09 & 0.15 & 0.16 & 0.20 \\
& & MAD & 0.03 & 0.10 & 0.17 & 0.19 & 0.24 \\
\bottomrule
\end{tabular}
\end{table}

\subsection{Robustness}
Robustness examines score stability across semantically equivalent messages with varying expressions, such as \textit{``Added support for Pine H64 board detection''} and \textit{``Incorporate PineH64 into board detection,''} which should receive comparable scores despite different lexical expressions.

\label{sec: robustness}
\subsubsection{Method}
This investigation examined LLM-based evaluators' ability to maintain scoring consistency across semantically equivalent commit messages. For each of the 400 developer-written messages, we constructed a triplet consisting of the original message and two paraphrased variants, with the variants generated by GPT-4 (generation prompts and implementation details are provided in the appendix \cite{appendix}). The first two authors manually verified semantic equivalence between the original messages and their variants, reaching a consensus on 342 triplets suitable for analysis. We employed LLM-based metrics three times for each message in these triplets and calculated the average as the final score. To assess consistency, we computed the standard deviation (std) and mean absolute difference (MAD) of final scores within each triplet, then averaged these metrics across all triplets. All LLM-based metric configurations followed the settings described in Sec. \ref{sec: reproducibility}.

\subsubsection{Results}
\label{sec: robustness_res}
Table \ref{tab:robustness} presents the average std and MAD for each model's evaluation of the 342 semantically equivalent commit messages. The results indicate moderate variations in LLM evaluations across different expressions of the same semantic commit message. Among the four evaluated models, GPT-4 shows the least variation, with an average std of 0.29 for the \textit{What} dimension and 0.25 for the \textit{Why} dimension (the maximum score is 4). Across all pairwise combinations within the 342 triplets, GPT-4's evaluation results show that 71.5\% of score variations were less than or equal to 0.5 and 93.8\% were less than or equal to 1 in the \textit{What} dimension; in the \textit{Why} dimension, 75.3\% of score variations were less than or equal to 0.5 and 97\% were less than or equal to 1. Considering that a one-point scoring difference is generally acceptable, GPT-4 demonstrates stability in 93.8\% of cases for \textit{What} and 97\% for \textit{Why} dimensions, suggesting acceptable robustness despite some variations.

\begin{table}[!h]
\centering
\small
\caption{Robustness analysis of different LLMs}
\label{tab:robustness}
\begin{tabular}{p{1.5cm}cccc}
\toprule
\multirow{2}{*}{\textbf{Model}} & \multicolumn{2}{c}{\textbf{What Dimension}} & \multicolumn{2}{c}{\textbf{Why Dimension}} \\
\cmidrule(lr){2-3}\cmidrule(lr){4-5}
& Std. & MAD & Std. & MAD \\
\midrule
GPT-4 & 0.29 & 0.36 & 0.25 & 0.31 \\
QwQ & 0.27 & 0.33 & 0.33 & 0.41 \\
Llama-3.3 & 0.28 & 0.33 & 0.44 & 0.54 \\
Qwen2.5 & 0.24 & 0.28 & 0.40 & 0.48 \\
\bottomrule
\end{tabular}
\end{table}

\subsection{Fairness}
\label{sec: bias}
Fairness investigates potential bias in LLM-based evaluators, particularly whether they exhibit bias toward LLM-generated messages compared to human-written ones.

\subsubsection{Method}
We used 342 triplets from Sec. \ref{sec: robustness}, each containing one developer-authored commit message and two semantically equivalent GPT-4-generated variants that deserved equivalent ratings on the \textit{What} and \textit{Why} dimensions. We analyzed the results from Sec. \ref{sec: robustness_res}, calculating the average scores assigned by LLMs to both developer-authored and LLM-generated messages to investigate potential scoring bias when evaluating semantically equivalent messages with varying expressions. Additionally, since our RQ1 utilized GPT-4 to evaluate human-authored commit messages and RQ2 also used GPT-4 to evaluate LLM-generated messages. In our dataset, both human-authored and LLM-generated messages are assessed by human developers, we identified 25 pairs (human-authored, LLM-generated) that received identical human developer scores on the \textit{What} dimension and 40 pairs with identical scores on the \textit{Why} dimension. We also report GPT-4's average scores for these pairs to investigate potential bias in its evaluation of human-authored versus LLM-generated messages.

\subsubsection{Results}

\begin{table}[!t]
\centering
\small
\caption{Evaluation scores comparison between human-authored and LLM-generated commit messages}
\label{tab:evaluation_comparison}
\begin{tabular}{lcccc}
\toprule
\multirow{2}{*}{\textbf{Model}} & \multicolumn{2}{c}{\textbf{Human-authored}} & \multicolumn{2}{c}{\textbf{LLM-generated}} \\
\cmidrule(lr){2-3}\cmidrule(lr){4-5}
& What & Why & What & Why \\
\midrule
GPT-4 & 2.74 & 1.42 & 2.81 $( \uparrow 0.08)$ & 1.49 $( \uparrow 0.08)$ \\
QwQ & 3.10 & 1.16 & 3.09 $( \downarrow 0.01)$ & 1.18 $( \uparrow 0.02)$ \\
Llama-3.3 & 3.18 & 1.87 & 3.22 $( \uparrow 0.04)$ & 1.98 $( \uparrow 0.11)$ \\
Qwen2.5 & 3.30 & 2.35 & 3.35 $( \uparrow 0.05)$ & 2.45 $( \uparrow 0.09)$ \\
\bottomrule
\end{tabular}
\end{table}

Table \ref{tab:evaluation_comparison} presents mean evaluation scores assigned by different LLMs to semantically equivalent messages. The results indicate consistent evaluations across human-written and LLM-generated messages, with marginal differences favoring LLM-generated messages (e.g., GPT-4 shows increases of 0.08 in both \textit{What} and \textit{Why} dimensions). Furthermore, in our dataset, among the 25 pairs receiving identical human ratings on the \textit{What} dimension, GPT-4 assigned an average score of 3.4 to human-authored messages and 3.72 to LLM-generated messages. For the 40 pairs with identical human ratings on the \textit{Why} dimension, GPT-4 assigned an equal average score of 0.85 to both message types. Overall, we did not observe substantial bias in LLM metrics, though GPT-4 scored LLM-generated messages 0.32 points higher on average, potentially indicating bias. However, due to the limited sample size, more comprehensive research on LLM fairness remains for future work.

\begin{summarybox}
{
\textbf{Summary for RQ3:} Due to the inherent non-deterministic behavior of LLMs, the LLM-based commit message evaluators exhibit some variations in reproducibility (identical inputs) and robustness (semantically equivalent but differently expressed inputs). However, the overall reproducibility and robustness remain within acceptable ranges. Our analysis revealed no substantial bias between human-written and LLM-generated commit messages, although more comprehensive research on LLM fairness is required.
}
\end{summarybox}

\section{Discussion and Conclusion}
\label{sec: discussions}

\subsection{Advantages and limitations of LLM-metrics}

The primary advantage of LLM-based evaluators lies in their higher correlation with human judgment compared to traditional automatic metrics. Our analysis shows that the correlation between LLM-based evaluators and human assessors approaches the inter-rater reliability among human annotators, suggesting near-human-level performance in commit message evaluation. Furthermore, LLM-based evaluation offers significant efficiency gains in both time and cost. While human evaluation required approximately six days for independent annotation of 800 commit messages in this study, LLM-based evaluation was completed within an hour. Cost analysis shows that using Chain-of-Thought and few-shot learning with unified evaluation criteria, GPT-4 evaluation averages US \$0.08 per example (calculated based on OpenAI API pricing\footnote{\url{https://openai.com/api/pricing/}} and the average tokens used in this study), which is substantially more cost-effective than employing an experienced developer for six days. This improved efficiency enables rapid feedback integration during model development or training.

Despite these advantages, LLM-based evaluators exhibit non-deterministic behavior, producing varying scores for identical messages or semantically equivalent expressions. Although these variations remain within acceptable bounds and can be mitigated by averaging scores across multiple evaluations, further research is needed to improve evaluator consistency and address potential biases. Future work could explore ensemble approaches combining multiple LLMs with voting mechanisms or incorporate additional context beyond code changes to enhance evaluation reliability.

\subsection{Threats to validity}
\label{sec: threats}
The threats to \textbf{external validity} primarily concern the generalizability of our findings. The non-deterministic nature of LLMs may introduce instability and, thus, potentially misleading results. To mitigate this threat, we conducted stability analysis examining the effects of different temperature settings and tested LLMs with semantically equivalent messages expressed in varying ways. The results indicate that LLM-based metrics indeed exhibit some instability, though remaining within acceptable ranges. Due to this inherent variability, LLM-based metrics are unsuitable for precise comparative analyses and should instead be viewed as approximations of human judgment. For instance, if Model1 achieves an LLM-metric score of 3.78 on the \textit{What} dimension while Model2 scores 3.71, this minor difference (relative to the standard deviation of 0.24 observed when using GPT-4 with temperature 0.25) should not be interpreted as definitive evidence of Model1's superiority, but rather suggests that both models perform at comparable levels on this dimension. Another threat is that our stability measurement in RQ3 explored reproducibility, robustness, and fairness from limited perspectives. There are other factors that may affect LLM metrics' reproducibility (e.g., software and hardware environment), and while our fairness evaluation did not observe significant bias when assessing semantically identical messages expressed differently, other comprehensive evaluations, such as assessing bias using semantically different examples that received consistent human ratings, should also be conducted (we only found 25 limited examples in this study). As a preliminary study, we only considered the significant factors, and a more comprehensive measurement of LLM metrics' stability would require more extensive future work.

The threats to \textbf{internal validity} primarily relate to potential biases in human annotation. To address this concern, we recruited volunteers with an average of more than five years of software development experience and conducted a three-stage annotation process. Each volunteer independently evaluated the commit messages, and the Spearman correlation coefficients between each pair of volunteers' ratings demonstrated strong inter-annotator agreement. Another threat is our 30\% threshold for high human ratings and low METEOR scores in our RQ2 thematic analysis. Higher thresholds would increase sample size but reduce accuracy by introducing more false positives. We chose 30\% to balance meaningful pattern detection with a manageable yet discriminative sample of 102 examples. 

\subsection{Conclusion}
\label{sec: conclusion}
This study presents a preliminary empirical investigation into utilizing Large Language Models (LLMs) for commit message quality evaluation. Our analysis demonstrates that incorporating Chain-of-Thought reasoning, few-shot learning, and unified evaluation criteria can enhance LLM-based evaluator performance. These evaluators substantially outperform all traditional automatic metrics in correlation with human judgment. Compared to human evaluation, LLM-based approaches offer superior efficiency in both time and cost. Integrating LLM-based evaluators in future commit message generation studies could facilitate better quality assessment and guide model development and training processes. While LLM-based evaluators show promising results, they present challenges regarding stability and potential biases. These limitations warrant careful consideration during implementation and necessitate further research for improvement.

\balance

\bibliographystyle{ACM-Reference-Format}
\bibliography{reference.bib}


\begin{thebibliography}{56}


\ifx \showCODEN    \undefined \def \showCODEN     #1{\unskip}     \fi
\ifx \showDOI      \undefined \def \showDOI       #1{#1}\fi
\ifx \showISBNx    \undefined \def \showISBNx     #1{\unskip}     \fi
\ifx \showISBNxiii \undefined \def \showISBNxiii  #1{\unskip}     \fi
\ifx \showISSN     \undefined \def \showISSN      #1{\unskip}     \fi
\ifx \showLCCN     \undefined \def \showLCCN      #1{\unskip}     \fi
\ifx \shownote     \undefined \def \shownote      #1{#1}          \fi
\ifx \showarticletitle \undefined \def \showarticletitle #1{#1}   \fi
\ifx \showURL      \undefined \def \showURL       {\relax}        \fi
\providecommand\bibfield[2]{#2}
\providecommand\bibinfo[2]{#2}
\providecommand\natexlab[1]{#1}
\providecommand\showeprint[2][]{arXiv:#2}

\bibitem[Achiam et~al\mbox{.}(2023)]%
        {achiam2023gpt}
\bibfield{author}{\bibinfo{person}{Josh Achiam}, \bibinfo{person}{Steven Adler}, \bibinfo{person}{Sandhini Agarwal}, \bibinfo{person}{Lama Ahmad}, \bibinfo{person}{Ilge Akkaya}, \bibinfo{person}{Florencia~Leoni Aleman}, \bibinfo{person}{Diogo Almeida}, \bibinfo{person}{Janko Altenschmidt}, \bibinfo{person}{Sam Altman}, \bibinfo{person}{Shyamal Anadkat}, {et~al\mbox{.}}} \bibinfo{year}{2023}\natexlab{}.
\newblock \showarticletitle{Gpt-4 technical report}.
\newblock \bibinfo{journal}{\emph{arXiv preprint arXiv:2303.08774}} (\bibinfo{year}{2023}).
\newblock


\bibitem[Author(s)(2025)]%
        {appendix}
\bibfield{author}{\bibinfo{person}{Anonymous Author(s)}.} \bibinfo{year}{2025}\natexlab{}.
\newblock \bibinfo{title}{Evaluating Generated Commit Messages with Large Language Models}.
\newblock \bibinfo{howpublished}{\url{https://anonymous.4open.science/r/4k8asu}}.
\newblock


\bibitem[Banerjee and Lavie(2005)]%
        {banerjee-lavie-2005-meteor}
\bibfield{author}{\bibinfo{person}{Satanjeev Banerjee} {and} \bibinfo{person}{Alon Lavie}.} \bibinfo{year}{2005}\natexlab{}.
\newblock \showarticletitle{{METEOR}: An Automatic Metric for {MT} Evaluation with Improved Correlation with Human Judgments}. In \bibinfo{booktitle}{\emph{Proceedings of the {ACL} Workshop on Intrinsic and Extrinsic Evaluation Measures for Machine Translation and/or Summarization}}, \bibfield{editor}{\bibinfo{person}{Jade Goldstein}, \bibinfo{person}{Alon Lavie}, \bibinfo{person}{Chin-Yew Lin}, {and} \bibinfo{person}{Clare Voss}} (Eds.). \bibinfo{publisher}{Association for Computational Linguistics}, \bibinfo{address}{Ann Arbor, Michigan}, \bibinfo{pages}{65--72}.
\newblock
\urldef\tempurl%
\url{https://aclanthology.org/W05-0909/}
\showURL{%
\tempurl}


\bibitem[Brown et~al\mbox{.}(2020)]%
        {brown2020language}
\bibfield{author}{\bibinfo{person}{Tom Brown}, \bibinfo{person}{Benjamin Mann}, \bibinfo{person}{Nick Ryder}, \bibinfo{person}{Melanie Subbiah}, \bibinfo{person}{Jared~D Kaplan}, \bibinfo{person}{Prafulla Dhariwal}, \bibinfo{person}{Arvind Neelakantan}, \bibinfo{person}{Pranav Shyam}, \bibinfo{person}{Girish Sastry}, \bibinfo{person}{Amanda Askell}, {et~al\mbox{.}}} \bibinfo{year}{2020}\natexlab{}.
\newblock \showarticletitle{Language models are few-shot learners}.
\newblock \bibinfo{journal}{\emph{Advances in neural information processing systems}}  \bibinfo{volume}{33} (\bibinfo{year}{2020}), \bibinfo{pages}{1877--1901}.
\newblock


\bibitem[Buse and Weimer(2010a)]%
        {2010Automatically}
\bibfield{author}{\bibinfo{person}{Rpl Buse} {and} \bibinfo{person}{W.~R. Weimer}.} \bibinfo{year}{2010}\natexlab{a}.
\newblock \showarticletitle{Automatically documenting program changes}. In \bibinfo{booktitle}{\emph{ASE 2010, 25th IEEE/ACM International Conference on Automated Software Engineering}}. \bibinfo{publisher}{{ACM}}, \bibinfo{pages}{33--42}.
\newblock


\bibitem[Buse and Weimer(2010b)]%
        {buse2010automatically}
\bibfield{author}{\bibinfo{person}{Raymond~PL Buse} {and} \bibinfo{person}{Westley~R Weimer}.} \bibinfo{year}{2010}\natexlab{b}.
\newblock \showarticletitle{Automatically documenting program changes}. In \bibinfo{booktitle}{\emph{Proceedings of the 25th IEEE/ACM international conference on automated software engineering}}. \bibinfo{pages}{33--42}.
\newblock


\bibitem[Chiang and yi~Lee(2023)]%
        {Chiang2023CanLL}
\bibfield{author}{\bibinfo{person}{Cheng-Han Chiang} {and} \bibinfo{person}{Hung yi Lee}.} \bibinfo{year}{2023}\natexlab{}.
\newblock \showarticletitle{Can Large Language Models Be an Alternative to Human Evaluations?}. In \bibinfo{booktitle}{\emph{Annual Meeting of the Association for Computational Linguistics}}.
\newblock
\urldef\tempurl%
\url{https://api.semanticscholar.org/CorpusID:258461287}
\showURL{%
\tempurl}


\bibitem[Cort{\'e}s-Coy et~al\mbox{.}(2014)]%
        {cortes2014automatically}
\bibfield{author}{\bibinfo{person}{Luis~Fernando Cort{\'e}s-Coy}, \bibinfo{person}{Mario Linares-V{\'a}squez}, \bibinfo{person}{Jairo Aponte}, {and} \bibinfo{person}{Denys Poshyvanyk}.} \bibinfo{year}{2014}\natexlab{}.
\newblock \showarticletitle{On automatically generating commit messages via summarization of source code changes}. In \bibinfo{booktitle}{\emph{IEEE 14th International Working Conference on Source Code Analysis and Manipulation}}. \bibinfo{pages}{275--284}.
\newblock


\bibitem[Cruzes and Dyba(2011)]%
        {cruzes2011recommended}
\bibfield{author}{\bibinfo{person}{Daniela~S Cruzes} {and} \bibinfo{person}{Tore Dyba}.} \bibinfo{year}{2011}\natexlab{}.
\newblock \showarticletitle{Recommended steps for thematic synthesis in software engineering}. In \bibinfo{booktitle}{\emph{2011 International Symposium on Empirical Software Engineering and Measurement}}. \bibinfo{publisher}{{IEEE} Computer Society}, \bibinfo{pages}{275--284}.
\newblock
\urldef\tempurl%
\url{https://doi.org/10.1109/ESEM.2011.36}
\showDOI{\tempurl}


\bibitem[Dong et~al\mbox{.}(2022)]%
        {dong2022fira}
\bibfield{author}{\bibinfo{person}{Jinhao Dong}, \bibinfo{person}{Yiling Lou}, \bibinfo{person}{Qihao Zhu}, \bibinfo{person}{Zeyu Sun}, \bibinfo{person}{Zhilin Li}, \bibinfo{person}{Wenjie Zhang}, {and} \bibinfo{person}{Dan Hao}.} \bibinfo{year}{2022}\natexlab{}.
\newblock \showarticletitle{FIRA: fine-grained graph-based code change representation for automated commit message generation}. In \bibinfo{booktitle}{\emph{Proceedings of the 44th International Conference on Software Engineering}}. \bibinfo{pages}{970--981}.
\newblock


\bibitem[Dubey et~al\mbox{.}(2024)]%
        {dubey2024llama}
\bibfield{author}{\bibinfo{person}{Abhimanyu Dubey}, \bibinfo{person}{Abhinav Jauhri}, \bibinfo{person}{Abhinav Pandey}, \bibinfo{person}{Abhishek Kadian}, \bibinfo{person}{Ahmad Al-Dahle}, \bibinfo{person}{Aiesha Letman}, \bibinfo{person}{Akhil Mathur}, \bibinfo{person}{Alan Schelten}, \bibinfo{person}{Amy Yang}, \bibinfo{person}{Angela Fan}, {et~al\mbox{.}}} \bibinfo{year}{2024}\natexlab{}.
\newblock \showarticletitle{The llama 3 herd of models}.
\newblock \bibinfo{journal}{\emph{arXiv preprint arXiv:2407.21783}} (\bibinfo{year}{2024}).
\newblock


\bibitem[Dyer et~al\mbox{.}(2013)]%
        {dyer2013boa}
\bibfield{author}{\bibinfo{person}{Robert Dyer}, \bibinfo{person}{Hoan~Anh Nguyen}, \bibinfo{person}{Hridesh Rajan}, {and} \bibinfo{person}{Tien~N Nguyen}.} \bibinfo{year}{2013}\natexlab{}.
\newblock \showarticletitle{Boa: A language and infrastructure for analyzing ultra-large-scale software repositories}. In \bibinfo{booktitle}{\emph{2013 35th International Conference on Software Engineering (ICSE)}}. IEEE, \bibinfo{pages}{422--431}.
\newblock


\bibitem[Guo et~al\mbox{.}(2025)]%
        {guo2025deepseek}
\bibfield{author}{\bibinfo{person}{Daya Guo}, \bibinfo{person}{Dejian Yang}, \bibinfo{person}{Haowei Zhang}, \bibinfo{person}{Junxiao Song}, \bibinfo{person}{Ruoyu Zhang}, \bibinfo{person}{Runxin Xu}, \bibinfo{person}{Qihao Zhu}, \bibinfo{person}{Shirong Ma}, \bibinfo{person}{Peiyi Wang}, \bibinfo{person}{Xiao Bi}, {et~al\mbox{.}}} \bibinfo{year}{2025}\natexlab{}.
\newblock \showarticletitle{Deepseek-r1: Incentivizing reasoning capability in llms via reinforcement learning}.
\newblock \bibinfo{journal}{\emph{arXiv preprint arXiv:2501.12948}} (\bibinfo{year}{2025}).
\newblock


\bibitem[Hoang et~al\mbox{.}(2020)]%
        {hoang2020cc2vec}
\bibfield{author}{\bibinfo{person}{Thong Hoang}, \bibinfo{person}{Hong~Jin Kang}, \bibinfo{person}{David Lo}, {and} \bibinfo{person}{Julia Lawall}.} \bibinfo{year}{2020}\natexlab{}.
\newblock \showarticletitle{CC2Vec: Distributed representations of code changes}. In \bibinfo{booktitle}{\emph{ACM/IEEE 42nd International Conference on Software Engineering}}. \bibinfo{pages}{518--529}.
\newblock


\bibitem[Hu et~al\mbox{.}(2022)]%
        {hu2022correlating}
\bibfield{author}{\bibinfo{person}{Xing Hu}, \bibinfo{person}{Qiuyuan Chen}, \bibinfo{person}{Haoye Wang}, \bibinfo{person}{Xin Xia}, \bibinfo{person}{David Lo}, {and} \bibinfo{person}{Thomas Zimmermann}.} \bibinfo{year}{2022}\natexlab{}.
\newblock \showarticletitle{Correlating automated and human evaluation of code documentation generation quality}.
\newblock \bibinfo{journal}{\emph{ACM Transactions on Software Engineering and Methodology (TOSEM)}} \bibinfo{volume}{31}, \bibinfo{number}{4} (\bibinfo{year}{2022}), \bibinfo{pages}{1--28}.
\newblock


\bibitem[huggingface(2025)]%
        {hf_evaluate}
\bibfield{author}{\bibinfo{person}{huggingface}.} \bibinfo{year}{2025}\natexlab{}.
\newblock \bibinfo{title}{Evaluate: A library for easily evaluating machine learning models and datasets}.
\newblock \bibinfo{howpublished}{\url{https://github.com/huggingface/evaluate}}.
\newblock


\bibitem[Jiang et~al\mbox{.}(2017)]%
        {jiang2017automatically}
\bibfield{author}{\bibinfo{person}{Siyuan Jiang}, \bibinfo{person}{Ameer Armaly}, {and} \bibinfo{person}{Collin McMillan}.} \bibinfo{year}{2017}\natexlab{}.
\newblock \showarticletitle{Automatically generating commit messages from diffs using neural machine translation}. In \bibinfo{booktitle}{\emph{2017 32nd IEEE/ACM International Conference on Automated Software Engineering}}. IEEE, \bibinfo{pages}{135--146}.
\newblock


\bibitem[Jiang et~al\mbox{.}(2024)]%
        {jiang2024self}
\bibfield{author}{\bibinfo{person}{Xue Jiang}, \bibinfo{person}{Yihong Dong}, \bibinfo{person}{Lecheng Wang}, \bibinfo{person}{Zheng Fang}, \bibinfo{person}{Qiwei Shang}, \bibinfo{person}{Ge Li}, \bibinfo{person}{Zhi Jin}, {and} \bibinfo{person}{Wenpin Jiao}.} \bibinfo{year}{2024}\natexlab{}.
\newblock \showarticletitle{Self-planning code generation with large language models}.
\newblock \bibinfo{journal}{\emph{ACM Transactions on Software Engineering and Methodology}} \bibinfo{volume}{33}, \bibinfo{number}{7} (\bibinfo{year}{2024}), \bibinfo{pages}{1--30}.
\newblock


\bibitem[Kendall(1938)]%
        {kendall1938new}
\bibfield{author}{\bibinfo{person}{Maurice~G Kendall}.} \bibinfo{year}{1938}\natexlab{}.
\newblock \showarticletitle{A new measure of rank correlation}.
\newblock \bibinfo{journal}{\emph{Biometrika}} \bibinfo{volume}{30}, \bibinfo{number}{1-2} (\bibinfo{year}{1938}), \bibinfo{pages}{81--93}.
\newblock


\bibitem[Kojima et~al\mbox{.}(2024)]%
        {cot}
\bibfield{author}{\bibinfo{person}{Takeshi Kojima}, \bibinfo{person}{Shixiang~Shane Gu}, \bibinfo{person}{Machel Reid}, \bibinfo{person}{Yutaka Matsuo}, {and} \bibinfo{person}{Yusuke Iwasawa}.} \bibinfo{year}{2024}\natexlab{}.
\newblock \showarticletitle{Large language models are zero-shot reasoners}. In \bibinfo{booktitle}{\emph{Proceedings of the 36th International Conference on Neural Information Processing Systems}} (New Orleans, LA, USA) \emph{(\bibinfo{series}{NIPS '22})}. \bibinfo{publisher}{Curran Associates Inc.}, \bibinfo{address}{Red Hook, NY, USA}, Article \bibinfo{articleno}{1613}, \bibinfo{numpages}{15}~pages.
\newblock
\showISBNx{9781713871088}


\bibitem[Kuckartz et~al\mbox{.}(2013)]%
        {kuckartz2013statistik}
\bibfield{author}{\bibinfo{person}{Udo Kuckartz}, \bibinfo{person}{Stefan R{\"a}diker}, \bibinfo{person}{Thomas Ebert}, {and} \bibinfo{person}{Julia Schehl}.} \bibinfo{year}{2013}\natexlab{}.
\newblock \bibinfo{booktitle}{\emph{Statistik: eine verst{\"a}ndliche Einf{\"u}hrung}}.
\newblock \bibinfo{publisher}{Springer-Verlag}.
\newblock


\bibitem[Kwon et~al\mbox{.}(2023)]%
        {kwon2023efficient}
\bibfield{author}{\bibinfo{person}{Woosuk Kwon}, \bibinfo{person}{Zhuohan Li}, \bibinfo{person}{Siyuan Zhuang}, \bibinfo{person}{Ying Sheng}, \bibinfo{person}{Lianmin Zheng}, \bibinfo{person}{Cody~Hao Yu}, \bibinfo{person}{Joseph Gonzalez}, \bibinfo{person}{Hao Zhang}, {and} \bibinfo{person}{Ion Stoica}.} \bibinfo{year}{2023}\natexlab{}.
\newblock \showarticletitle{Efficient memory management for large language model serving with pagedattention}. In \bibinfo{booktitle}{\emph{Proceedings of the 29th Symposium on Operating Systems Principles}}. \bibinfo{pages}{611--626}.
\newblock


\bibitem[Li and Ahmed(2023)]%
        {li2023commit}
\bibfield{author}{\bibinfo{person}{Jiawei Li} {and} \bibinfo{person}{Iftekhar Ahmed}.} \bibinfo{year}{2023}\natexlab{}.
\newblock \showarticletitle{Commit message matters: Investigating impact and evolution of commit message quality}. In \bibinfo{booktitle}{\emph{2023 IEEE/ACM 45th International Conference on Software Engineering (ICSE)}}. IEEE, \bibinfo{pages}{806--817}.
\newblock


\bibitem[Li et~al\mbox{.}(2024)]%
        {diffisno}
\bibfield{author}{\bibinfo{person}{Jiawei Li}, \bibinfo{person}{David Farag\'{o}}, \bibinfo{person}{Christian Petrov}, {and} \bibinfo{person}{Iftekhar Ahmed}.} \bibinfo{year}{2024}\natexlab{}.
\newblock \showarticletitle{Only diff Is Not Enough: Generating Commit Messages Leveraging Reasoning and Action of Large Language Model}.
\newblock \bibinfo{journal}{\emph{Proc. ACM Softw. Eng.}} \bibinfo{volume}{1}, \bibinfo{number}{FSE}, Article \bibinfo{articleno}{34} (\bibinfo{date}{July} \bibinfo{year}{2024}), \bibinfo{numpages}{22}~pages.
\newblock
\urldef\tempurl%
\url{https://doi.org/10.1145/3643760}
\showDOI{\tempurl}


\bibitem[Lin and Och(2004)]%
        {lin2004rouge}
\bibfield{author}{\bibinfo{person}{Chin-Yew Lin} {and} \bibinfo{person}{Franz~Josef Och}.} \bibinfo{year}{2004}\natexlab{}.
\newblock \showarticletitle{Automatic evaluation of machine translation quality using longest common subsequence and skip-bigram statistics}. In \bibinfo{booktitle}{\emph{Proceedings of the 42nd annual meeting of the association for computational linguistics (ACL-04)}}. \bibinfo{pages}{605--612}.
\newblock


\bibitem[Liu et~al\mbox{.}(2024)]%
        {liu2024deepseek}
\bibfield{author}{\bibinfo{person}{Aixin Liu}, \bibinfo{person}{Bei Feng}, \bibinfo{person}{Bing Xue}, \bibinfo{person}{Bingxuan Wang}, \bibinfo{person}{Bochao Wu}, \bibinfo{person}{Chengda Lu}, \bibinfo{person}{Chenggang Zhao}, \bibinfo{person}{Chengqi Deng}, \bibinfo{person}{Chenyu Zhang}, \bibinfo{person}{Chong Ruan}, {et~al\mbox{.}}} \bibinfo{year}{2024}\natexlab{}.
\newblock \showarticletitle{DeepSeek-V3 Technical Report}.
\newblock \bibinfo{journal}{\emph{arXiv preprint arXiv:2412.19437}} (\bibinfo{year}{2024}).
\newblock


\bibitem[Liu et~al\mbox{.}(2019b)]%
        {liu2019generating}
\bibfield{author}{\bibinfo{person}{Qin Liu}, \bibinfo{person}{Zihe Liu}, \bibinfo{person}{Hongming Zhu}, \bibinfo{person}{Hongfei Fan}, \bibinfo{person}{Bowen Du}, {and} \bibinfo{person}{Yu Qian}.} \bibinfo{year}{2019}\natexlab{b}.
\newblock \showarticletitle{Generating commit messages from diffs using pointer-generator network}. In \bibinfo{booktitle}{\emph{2019 IEEE/ACM 16th International Conference on Mining Software Repositories}}. IEEE, \bibinfo{pages}{299--309}.
\newblock


\bibitem[Liu et~al\mbox{.}(2019a)]%
        {liu2020atom}
\bibfield{author}{\bibinfo{person}{Shangqing Liu}, \bibinfo{person}{Cuiyun Gao}, \bibinfo{person}{Sen Chen}, \bibinfo{person}{Lun~Yiu Nie}, {and} \bibinfo{person}{Yang Liu}.} \bibinfo{year}{2019}\natexlab{a}.
\newblock \showarticletitle{{ATOM:} Commit Message Generation Based on Abstract Syntax Tree and Hybrid Ranking}.
\newblock \bibinfo{journal}{\emph{CoRR}}  \bibinfo{volume}{abs/1912.02972} (\bibinfo{year}{2019}).
\newblock
\showeprint[arxiv]{1912.02972}
\urldef\tempurl%
\url{http://arxiv.org/abs/1912.02972}
\showURL{%
\tempurl}


\bibitem[Liu et~al\mbox{.}(2023)]%
        {liu-etal-2023-g}
\bibfield{author}{\bibinfo{person}{Yang Liu}, \bibinfo{person}{Dan Iter}, \bibinfo{person}{Yichong Xu}, \bibinfo{person}{Shuohang Wang}, \bibinfo{person}{Ruochen Xu}, {and} \bibinfo{person}{Chenguang Zhu}.} \bibinfo{year}{2023}\natexlab{}.
\newblock \showarticletitle{{G}-Eval: {NLG} Evaluation using Gpt-4 with Better Human Alignment}. In \bibinfo{booktitle}{\emph{Proceedings of the 2023 Conference on Empirical Methods in Natural Language Processing}}, \bibfield{editor}{\bibinfo{person}{Houda Bouamor}, \bibinfo{person}{Juan Pino}, {and} \bibinfo{person}{Kalika Bali}} (Eds.). \bibinfo{publisher}{Association for Computational Linguistics}, \bibinfo{address}{Singapore}, \bibinfo{pages}{2511--2522}.
\newblock
\urldef\tempurl%
\url{https://doi.org/10.18653/v1/2023.emnlp-main.153}
\showDOI{\tempurl}


\bibitem[Liu et~al\mbox{.}(2018)]%
        {liu2018neural}
\bibfield{author}{\bibinfo{person}{Zhongxin Liu}, \bibinfo{person}{Xin Xia}, \bibinfo{person}{Ahmed~E. Hassan}, \bibinfo{person}{David Lo}, \bibinfo{person}{Zhenchang Xing}, {and} \bibinfo{person}{Xinyu Wang}.} \bibinfo{year}{2018}\natexlab{}.
\newblock \showarticletitle{Neural-machine-translation-based commit message generation: how far are we?}. In \bibinfo{booktitle}{\emph{33rd {ACM/IEEE} International Conference on Automated Software Engineering, {ASE} 2018, Montpellier, France, September 3-7, 2018}}. \bibinfo{publisher}{{ACM}}, \bibinfo{pages}{373--384}.
\newblock


\bibitem[Loyola et~al\mbox{.}(2018)]%
        {loyola2018content}
\bibfield{author}{\bibinfo{person}{Pablo Loyola}, \bibinfo{person}{Edison Marrese-Taylor}, \bibinfo{person}{Jorge Balazs}, \bibinfo{person}{Yutaka Matsuo}, {and} \bibinfo{person}{Fumiko Satoh}.} \bibinfo{year}{2018}\natexlab{}.
\newblock \showarticletitle{Content aware source code change description generation}. In \bibinfo{booktitle}{\emph{11th International Conference on Natural Language Generation}}. \bibinfo{pages}{119--128}.
\newblock


\bibitem[Loyola et~al\mbox{.}(2017)]%
        {Loyola_2017}
\bibfield{author}{\bibinfo{person}{Pablo Loyola}, \bibinfo{person}{Edison Marrese-Taylor}, {and} \bibinfo{person}{Yutaka Matsuo}.} \bibinfo{year}{2017}\natexlab{}.
\newblock \showarticletitle{A Neural Architecture for Generating Natural Language Descriptions from Source Code Changes}. In \bibinfo{booktitle}{\emph{55th Annual Meeting of the Association for Computational Linguistics (Volume 2: Short Papers)}}. \bibinfo{publisher}{Association for Computational Linguistics}, \bibinfo{address}{Vancouver, Canada}, \bibinfo{pages}{287--292}.
\newblock


\bibitem[Nie et~al\mbox{.}(2021)]%
        {nie2021coregen}
\bibfield{author}{\bibinfo{person}{Lun~Yiu Nie}, \bibinfo{person}{Cuiyun Gao}, \bibinfo{person}{Zhicong Zhong}, \bibinfo{person}{Wai Lam}, \bibinfo{person}{Yang Liu}, {and} \bibinfo{person}{Zenglin Xu}.} \bibinfo{year}{2021}\natexlab{}.
\newblock \showarticletitle{Coregen: Contextualized code representation learning for commit message generation}.
\newblock \bibinfo{journal}{\emph{Neurocomputing}}  \bibinfo{volume}{459} (\bibinfo{year}{2021}), \bibinfo{pages}{97--107}.
\newblock


\bibitem[Papineni et~al\mbox{.}(2002)]%
        {papineni2002bleu}
\bibfield{author}{\bibinfo{person}{Kishore Papineni}, \bibinfo{person}{Salim Roukos}, \bibinfo{person}{Todd Ward}, {and} \bibinfo{person}{Wei-Jing Zhu}.} \bibinfo{year}{2002}\natexlab{}.
\newblock \showarticletitle{Bleu: a method for automatic evaluation of machine translation}. In \bibinfo{booktitle}{\emph{Proceedings of the 40th annual meeting of the Association for Computational Linguistics}}. \bibinfo{pages}{311--318}.
\newblock


\bibitem[Reimers(2019)]%
        {reimers2019sentence}
\bibfield{author}{\bibinfo{person}{N Reimers}.} \bibinfo{year}{2019}\natexlab{}.
\newblock \showarticletitle{Sentence-BERT: Sentence Embeddings using Siamese BERT-Networks}.
\newblock \bibinfo{journal}{\emph{arXiv preprint arXiv:1908.10084}} (\bibinfo{year}{2019}).
\newblock


\bibitem[Schall et~al\mbox{.}(2024)]%
        {schall2024commitbench}
\bibfield{author}{\bibinfo{person}{Maxmilian Schall}, \bibinfo{person}{Tamara Czinczoll}, {and} \bibinfo{person}{Gerard De~Melo}.} \bibinfo{year}{2024}\natexlab{}.
\newblock \showarticletitle{CommitBench: A benchmark for commit message generation}. In \bibinfo{booktitle}{\emph{2024 IEEE International Conference on Software Analysis, Evolution and Reengineering (SANER)}}. IEEE, \bibinfo{pages}{728--739}.
\newblock


\bibitem[Shen et~al\mbox{.}(2016)]%
        {2016shen}
\bibfield{author}{\bibinfo{person}{Jinfeng Shen}, \bibinfo{person}{Xiaobing Sun}, \bibinfo{person}{Bin Li}, \bibinfo{person}{Hui Yang}, {and} \bibinfo{person}{Jiajun Hu}.} \bibinfo{year}{2016}\natexlab{}.
\newblock \showarticletitle{On automatic summarization of what and why information in source code changes}. In \bibinfo{booktitle}{\emph{2016 IEEE 40th Annual Computer Software and Applications Conference}}, Vol.~\bibinfo{volume}{1}. IEEE, \bibinfo{pages}{103--112}.
\newblock


\bibitem[Shi et~al\mbox{.}(2022)]%
        {shi2022-race}
\bibfield{author}{\bibinfo{person}{Ensheng Shi}, \bibinfo{person}{Yanlin Wang}, \bibinfo{person}{Wei Tao}, \bibinfo{person}{Lun Du}, \bibinfo{person}{Hongyu Zhang}, \bibinfo{person}{Shi Han}, \bibinfo{person}{Dongmei Zhang}, {and} \bibinfo{person}{Hongbin Sun}.} \bibinfo{year}{2022}\natexlab{}.
\newblock \showarticletitle{{RACE}: Retrieval-augmented Commit Message Generation}. In \bibinfo{booktitle}{\emph{Proceedings of the 2022 Conference on Empirical Methods in Natural Language Processing}}, \bibfield{editor}{\bibinfo{person}{Yoav Goldberg}, \bibinfo{person}{Zornitsa Kozareva}, {and} \bibinfo{person}{Yue Zhang}} (Eds.). \bibinfo{publisher}{Association for Computational Linguistics}, \bibinfo{address}{Abu Dhabi, United Arab Emirates}, \bibinfo{pages}{5520--5530}.
\newblock
\urldef\tempurl%
\url{https://doi.org/10.18653/v1/2022.emnlp-main.372}
\showDOI{\tempurl}


\bibitem[Spearman(1961)]%
        {spearman1961proof}
\bibfield{author}{\bibinfo{person}{Charles Spearman}.} \bibinfo{year}{1961}\natexlab{}.
\newblock \showarticletitle{The proof and measurement of association between two things.}
\newblock  (\bibinfo{year}{1961}).
\newblock


\bibitem[Team(2025)]%
        {qwq}
\bibfield{author}{\bibinfo{person}{Qwen Team}.} \bibinfo{year}{2025}\natexlab{}.
\newblock \bibinfo{title}{QwQ-32B: Embracing the Power of Reinforcement Learning}.
\newblock \bibinfo{howpublished}{\url{https://qwenlm.github.io/blog/qwq-32b/}}.
\newblock


\bibitem[Tian et~al\mbox{.}(2022)]%
        {tian2022makes}
\bibfield{author}{\bibinfo{person}{Yingchen Tian}, \bibinfo{person}{Yuxia Zhang}, \bibinfo{person}{Klaas-Jan Stol}, \bibinfo{person}{Lin Jiang}, {and} \bibinfo{person}{Hui Liu}.} \bibinfo{year}{2022}\natexlab{}.
\newblock \showarticletitle{What makes a good commit message?}. In \bibinfo{booktitle}{\emph{Proceedings of the 44th International Conference on Software Engineering}}. \bibinfo{pages}{2389--2401}.
\newblock


\bibitem[Touvron et~al\mbox{.}(2023)]%
        {touvron2023llama}
\bibfield{author}{\bibinfo{person}{Hugo Touvron}, \bibinfo{person}{Thibaut Lavril}, \bibinfo{person}{Gautier Izacard}, \bibinfo{person}{Xavier Martinet}, \bibinfo{person}{Marie-Anne Lachaux}, \bibinfo{person}{Timoth{\'e}e Lacroix}, \bibinfo{person}{Baptiste Rozi{\`e}re}, \bibinfo{person}{Naman Goyal}, \bibinfo{person}{Eric Hambro}, \bibinfo{person}{Faisal Azhar}, {et~al\mbox{.}}} \bibinfo{year}{2023}\natexlab{}.
\newblock \showarticletitle{Llama: Open and efficient foundation language models}.
\newblock \bibinfo{journal}{\emph{arXiv preprint arXiv:2302.13971}} (\bibinfo{year}{2023}).
\newblock


\bibitem[Vaswani et~al\mbox{.}(2017)]%
        {vaswani2017attention}
\bibfield{author}{\bibinfo{person}{Ashish Vaswani}, \bibinfo{person}{Noam Shazeer}, \bibinfo{person}{Niki Parmar}, \bibinfo{person}{Jakob Uszkoreit}, \bibinfo{person}{Llion Jones}, \bibinfo{person}{Aidan~N Gomez}, \bibinfo{person}{{\L}ukasz Kaiser}, {and} \bibinfo{person}{Illia Polosukhin}.} \bibinfo{year}{2017}\natexlab{}.
\newblock \showarticletitle{Attention is all you need}.
\newblock \bibinfo{journal}{\emph{Advances in neural information processing systems}}  \bibinfo{volume}{30} (\bibinfo{year}{2017}).
\newblock


\bibitem[Vedantam et~al\mbox{.}(2015)]%
        {vedantam2015cider}
\bibfield{author}{\bibinfo{person}{Ramakrishna Vedantam}, \bibinfo{person}{C Lawrence~Zitnick}, {and} \bibinfo{person}{Devi Parikh}.} \bibinfo{year}{2015}\natexlab{}.
\newblock \showarticletitle{Cider: Consensus-based image description evaluation}. In \bibinfo{booktitle}{\emph{Proceedings of the IEEE conference on computer vision and pattern recognition}}. \bibinfo{pages}{4566--4575}.
\newblock


\bibitem[Wang et~al\mbox{.}({[n.\,d.]})]%
        {wanghard}
\bibfield{author}{\bibinfo{person}{Guoqing Wang}, \bibinfo{person}{Zeyu Sun}, \bibinfo{person}{Jinhao Dong}, \bibinfo{person}{Yuxia Zhang}, \bibinfo{person}{Mingxuan Zhu}, \bibinfo{person}{Qingyuan Liang}, {and} \bibinfo{person}{Dan Hao}.} \bibinfo{year}{[n.\,d.]}\natexlab{}.
\newblock \showarticletitle{Is It Hard to Generate Holistic Commit Message?}
\newblock \bibinfo{journal}{\emph{ACM Transactions on Software Engineering and Methodology}} (\bibinfo{year}{[n.\,d.]}).
\newblock


\bibitem[Wang et~al\mbox{.}(2024)]%
        {10.1145/3695996}
\bibfield{author}{\bibinfo{person}{Guoqing Wang}, \bibinfo{person}{Zeyu Sun}, \bibinfo{person}{Jinhao Dong}, \bibinfo{person}{Yuxia Zhang}, \bibinfo{person}{Mingxuan Zhu}, \bibinfo{person}{Qingyuan Liang}, {and} \bibinfo{person}{Dan Hao}.} \bibinfo{year}{2024}\natexlab{}.
\newblock \showarticletitle{Is It Hard to Generate Holistic Commit Message?}
\newblock \bibinfo{journal}{\emph{ACM Trans. Softw. Eng. Methodol.}} (\bibinfo{date}{Sept.} \bibinfo{year}{2024}).
\newblock
\showISSN{1049-331X}
\urldef\tempurl%
\url{https://doi.org/10.1145/3695996}
\showDOI{\tempurl}
\newblock
\shownote{Just Accepted}.


\bibitem[Wang et~al\mbox{.}(2021)]%
        {wang2021context}
\bibfield{author}{\bibinfo{person}{Haoye Wang}, \bibinfo{person}{Xin Xia}, \bibinfo{person}{David Lo}, \bibinfo{person}{Qiang He}, \bibinfo{person}{Xinyu Wang}, {and} \bibinfo{person}{John Grundy}.} \bibinfo{year}{2021}\natexlab{}.
\newblock \showarticletitle{Context-Aware Retrieval-Based Deep Commit Message Generation}.
\newblock \bibinfo{journal}{\emph{ACM Transactions on Software Engineering and Methodology}} \bibinfo{volume}{30}, \bibinfo{number}{4}, Article \bibinfo{articleno}{56} (\bibinfo{year}{2021}), \bibinfo{numpages}{30}~pages.
\newblock
\showISSN{1049-331X}


\bibitem[Wei et~al\mbox{.}(2022)]%
        {wei2022chain}
\bibfield{author}{\bibinfo{person}{Jason Wei}, \bibinfo{person}{Xuezhi Wang}, \bibinfo{person}{Dale Schuurmans}, \bibinfo{person}{Maarten Bosma}, \bibinfo{person}{Fei Xia}, \bibinfo{person}{Ed Chi}, \bibinfo{person}{Quoc~V Le}, \bibinfo{person}{Denny Zhou}, {et~al\mbox{.}}} \bibinfo{year}{2022}\natexlab{}.
\newblock \showarticletitle{Chain-of-thought prompting elicits reasoning in large language models}.
\newblock \bibinfo{journal}{\emph{Advances in neural information processing systems}}  \bibinfo{volume}{35} (\bibinfo{year}{2022}), \bibinfo{pages}{24824--24837}.
\newblock


\bibitem[Xu et~al\mbox{.}(2019)]%
        {xu2019commit}
\bibfield{author}{\bibinfo{person}{Shengbin Xu}, \bibinfo{person}{Yuan Yao}, \bibinfo{person}{Feng Xu}, \bibinfo{person}{Tianxiao Gu}, \bibinfo{person}{Hanghang Tong}, {and} \bibinfo{person}{Jian Lu}.} \bibinfo{year}{2019}\natexlab{}.
\newblock \showarticletitle{Commit Message Generation for Source Code Changes}. In \bibinfo{booktitle}{\emph{Twenty-Eighth International Joint Conference on Artificial Intelligence}}. \bibinfo{pages}{3975--3981}.
\newblock


\bibitem[Xue et~al\mbox{.}(2024)]%
        {10713474}
\bibfield{author}{\bibinfo{person}{Pengyu Xue}, \bibinfo{person}{Linhao Wu}, \bibinfo{person}{Zhongxing Yu}, \bibinfo{person}{Zhi Jin}, \bibinfo{person}{Zhen Yang}, \bibinfo{person}{Xinyi Li}, \bibinfo{person}{Zhenyu Yang}, {and} \bibinfo{person}{Yue Tan}.} \bibinfo{year}{2024}\natexlab{}.
\newblock \showarticletitle{Automated Commit Message Generation With Large Language Models: An Empirical Study and Beyond}.
\newblock \bibinfo{journal}{\emph{IEEE Transactions on Software Engineering}} \bibinfo{volume}{50}, \bibinfo{number}{12} (\bibinfo{year}{2024}), \bibinfo{pages}{3208--3224}.
\newblock
\urldef\tempurl%
\url{https://doi.org/10.1109/TSE.2024.3478317}
\showDOI{\tempurl}


\bibitem[Yang et~al\mbox{.}(2024)]%
        {yang2024qwen2}
\bibfield{author}{\bibinfo{person}{An Yang}, \bibinfo{person}{Baosong Yang}, \bibinfo{person}{Beichen Zhang}, \bibinfo{person}{Binyuan Hui}, \bibinfo{person}{Bo Zheng}, \bibinfo{person}{Bowen Yu}, \bibinfo{person}{Chengyuan Li}, \bibinfo{person}{Dayiheng Liu}, \bibinfo{person}{Fei Huang}, \bibinfo{person}{Haoran Wei}, {et~al\mbox{.}}} \bibinfo{year}{2024}\natexlab{}.
\newblock \showarticletitle{Qwen2. 5 Technical Report}.
\newblock \bibinfo{journal}{\emph{arXiv preprint arXiv:2412.15115}} (\bibinfo{year}{2024}).
\newblock


\bibitem[Zhang et~al\mbox{.}(2024b)]%
        {zhang2024using}
\bibfield{author}{\bibinfo{person}{Linghao Zhang}, \bibinfo{person}{Jingshu Zhao}, \bibinfo{person}{Chong Wang}, {and} \bibinfo{person}{Peng Liang}.} \bibinfo{year}{2024}\natexlab{b}.
\newblock \showarticletitle{Using Large Language Models for Commit Message Generation: A Preliminary Study}.
\newblock \bibinfo{journal}{\emph{arXiv preprint arXiv:2401.05926}} (\bibinfo{year}{2024}).
\newblock


\bibitem[Zhang et~al\mbox{.}(2019)]%
        {zhang2019bertscore}
\bibfield{author}{\bibinfo{person}{Tianyi Zhang}, \bibinfo{person}{Varsha Kishore}, \bibinfo{person}{Felix Wu}, \bibinfo{person}{Kilian~Q Weinberger}, {and} \bibinfo{person}{Yoav Artzi}.} \bibinfo{year}{2019}\natexlab{}.
\newblock \showarticletitle{Bertscore: Evaluating text generation with bert}.
\newblock \bibinfo{journal}{\emph{arXiv preprint arXiv:1904.09675}} (\bibinfo{year}{2019}).
\newblock


\bibitem[Zhang et~al\mbox{.}(2020)]%
        {bert-score}
\bibfield{author}{\bibinfo{person}{Tianyi Zhang}, \bibinfo{person}{Varsha Kishore}, \bibinfo{person}{Felix Wu}, \bibinfo{person}{Kilian~Q. Weinberger}, {and} \bibinfo{person}{Yoav Artzi}.} \bibinfo{year}{2020}\natexlab{}.
\newblock \showarticletitle{BERTScore: Evaluating Text Generation with BERT}. In \bibinfo{booktitle}{\emph{International Conference on Learning Representations}}.
\newblock
\urldef\tempurl%
\url{https://openreview.net/forum?id=SkeHuCVFDr}
\showURL{%
\tempurl}


\bibitem[Zhang et~al\mbox{.}(2024a)]%
        {zhang2024automatic}
\bibfield{author}{\bibinfo{person}{Yuxia Zhang}, \bibinfo{person}{Zhiqing Qiu}, \bibinfo{person}{Klaas-Jan Stol}, \bibinfo{person}{Wenhui Zhu}, \bibinfo{person}{Jiaxin Zhu}, \bibinfo{person}{Yingchen Tian}, {and} \bibinfo{person}{Hui Liu}.} \bibinfo{year}{2024}\natexlab{a}.
\newblock \showarticletitle{Automatic commit message generation: A critical review and directions for future work}.
\newblock \bibinfo{journal}{\emph{IEEE Transactions on Software Engineering}} (\bibinfo{year}{2024}).
\newblock


\bibitem[Zhao et~al\mbox{.}(2023)]%
        {zhao2023survey}
\bibfield{author}{\bibinfo{person}{Wayne~Xin Zhao}, \bibinfo{person}{Kun Zhou}, \bibinfo{person}{Junyi Li}, \bibinfo{person}{Tianyi Tang}, \bibinfo{person}{Xiaolei Wang}, \bibinfo{person}{Yupeng Hou}, \bibinfo{person}{Yingqian Min}, \bibinfo{person}{Beichen Zhang}, \bibinfo{person}{Junjie Zhang}, \bibinfo{person}{Zican Dong}, {et~al\mbox{.}}} \bibinfo{year}{2023}\natexlab{}.
\newblock \showarticletitle{A survey of large language models}.
\newblock \bibinfo{journal}{\emph{arXiv preprint arXiv:2303.18223}} (\bibinfo{year}{2023}).
\newblock


\end{thebibliography}

\end{document}